\documentclass[aps,reprint,showpacs,amsmath,amssymb,prc,floatfix,nofootinbib]{revtex4-1}
\usepackage{color}
\usepackage{graphicx}% Include figure files
\usepackage{dcolumn}% Align table columns on decimal point
\usepackage{bm}% bold math
\usepackage{rotating}
\usepackage{float}
\usepackage{xcolor}
\usepackage{natbib}
\usepackage{color}
\usepackage{times}
\usepackage{inputenc}
\usepackage[colorlinks,citecolor=blue,linkcolor=red,anchorcolor=blue,filecolor=blue,urlcolor=blue]{hyperref}
\usepackage{comment}
\begin{document}
\title{Application of coherent density fluctuation model to study the nuclear matter properties of finite nuclei within relativistic mean-field formalism}

\author{Ankit Kumar$^{1,2}$}
\email{ankit.k@iopb.res.in}
\author{H. C. Das$^{1,2}$}
\author{Manpreet Kaur$^{1,2}$}
\author{M. Bhuyan$^{3,4}$}
\email{bunuphy@yahoo.com}
\author{S. K. Patra$^{1,2}$}
%\email{patra@iopb.res.in}
%%%%%%%%%%%%%%%%%%%%%%%%%%%%%%%%%%%%%%

\affiliation{\it $^{1}$Institute of Physics, Sachivalya Marg, Bhubaneswar-751005, India}
\affiliation{\it $^{2}$Homi Bhabha National Institute, Training School Complex,
Anushakti Nagar, Mumbai 400094, India}
\affiliation{\it $^{3}$Department of Physics, Faculty of Science, University of Malaya, Kuala Lumpur 50603, Malaysia}
\affiliation{\it $^{4}$Institute of Research and Development, Duy Tan University, Da Nang 550000, Vietnam}
\date{\today}

\begin{abstract}
We obtained a density-dependent analytical expression of binding energy per nucleon for different neutron-proton asymmetry of the nuclear matter (NM) with a polynomial fitting, which manifests the results of effective field theory motivated relativistic mean-field (E-RMF) model. This expression has the edge over the Br$\ddot{u}$ckner energy density functional [Phys. Rev. {\bf 171}, 1188 (1968)] since it resolves the Coster-Band problem. The NM parameters like incompressibility, neutron pressure, symmetry energy, and its derivatives are calculated using the acquired expression of energy per nucleon. Further, the weight function calculated by E-RMF densities are folded with calculated NM parameters within coherent density fluctuation model to find the properties of closed/semi-closed-shell even-even $^{16}$O, $^{40}$Ca, $^{48}$Ca, $^{56}$Ni, $^{90}$Zr, $^{116}$Sn, and $^{208}$Pb nuclei. The values obtained for the neutron pressure $P^{A}$, symmetry energy $S^{A}$ and its derivative $L_{sym}^A$ known as slope parameter, lie within a narrow domain whereas there is a large variation in isoscalar incompressibility $K^{A}$ and surface incompressibility $K_{sym}^{A}$ while moving from light to heavy nuclei. The sizable variation in $K^{A}$ and $K_{sym}^{A}$ for light and heavy nuclei depicts their structural dependence due to the peculiar density distribution of each nucleus. A comparison of surface quantities calculated in the present work has also been made with ones obtained via Br$\ddot{u}$ckner energy density functional.
\end{abstract}
\maketitle
%%%%%%%%%%%%%%%%

\section{Introduction}\label{sec1}
The correlations among the nuclear matter (NM) and finite nuclei in terms of symmetry energy and its coefficients play a crucial role not only in nuclear physics but also in astrophysics. The isospin dependence of symmetry energy imparts information about the isovector component of the nuclear interaction, which is directly connected with the skin thickness of the nuclei. Eventually, different studies such as the island of stability of exotic nuclei, the dynamics of heavy-ion collisions, dipole polarizability, properties of neutron star (NS), core-collapse of massive compact stars and the nucleosynthesis process through neutrino convection at high-density region upon the symmetry energy and its coefficients \cite{latimer2000,latimer01,steiner05,li08,lynch09,tsang12,horo14,gad16,bhu18,kaurnpa, kaurjpg, ijmpe}. Therefore, it is indispensable to determine the symmetry energy and its coefficients for finite nuclei. Recently, many efforts have been made on theoretical as well as experimental fronts to probe the isospin dependence of symmetry energy and its coefficients, which is an ultimate bridge between finite nuclei and infinite NM \cite{steiner05,tsang12,horo14,gad16,bhu18,kaurnpa,kaurjpg,ijmpe}. Moreover, in some recent works, it is established that the kink in the symmetry energy of finite nuclei over the isotopic chain infers the appearance of shell/sub-shell closures \cite{gad16,bhu18,kaurnpa,kaurjpg}.

In recent works \cite{bhu18,abdul19}, the Br$\ddot{u}$ckner energy density functional \cite{bruk68,bruk69} has been used within the coherent density fluctuation model (CDFM)  to calculate the properties of nuclei. This functional had been fitted to the kinetic and potential energy parts to get the analytical expression of binding energy per particle E/A in the local density approximation (LDA) of the Thomas-Fermi approach. It is relevant to point out that Br$\ddot{u}$ckner energy functional does not respect the ``Coester-Band'' i.e. in Thomas-Fermi approach NM saturates at $\rho\sim 0.2$ fm$^{-3}$ instead of $\rho\sim 0.15$ fm$^{-3}$ \cite{coester,brockmann}. To have some meaningful correlations while extrapolating to higher densities, the nuclear equation of state (EOS) must satisfy the nuclear saturation properties. To address this problem, we have fitted the NM saturation plots for different values of asymmetry parameter, obtained using effective field theory motivated relativistic mean-field (E-RMF) model \cite{kumar18,kumar17} with standard NL3 and recently developed G3 parameter sets for the first time. The different NM parameters such as incompressibility, symmetry energy, and its derivatives are obtained employing the derived expression of E-RMF density functional. Subsequently, theses NM parameters are used along with E-RMF densities within the CDFM to find the corresponding quantities for magic/semi-magic $^{16}$O, $^{40}$Ca, $^{48}$Ca, $^{56}$Ni, $^{90}$Zr, $^{116}$Sn, and $^{208}$Pb nuclei.\\
\newline
The paper is organized as follows: The E-RMF approach and the fitting procedure to get the analytical polynomial expression of $E/A$ is discussed in Sec. \ref{theory}. The CDFM is also discussed in this section. Sec. \ref{results} is assigned to the discussion of the results obtained from the calculations. A brief summary and conclusions are presented in Sec. \ref{summary}.
%%%%%%

%
\section{Effective field theory motivated relativistic mean field (E-RMF) Model}
\label{theory}
In this section, we briefly describe the formalism of recently developed E-RMF model. The E-RMF Lagrangian density is constructed by taking the interactions of isoscalar (scalar $\sigma$, vector $\omega$) and isovector (scalar $\delta$, vector $\rho$) mesons with nucleons and among themselves. The E-RMF Lagrangian is discussed in Refs. \cite{kumar18,kumar17,frun96,frun97}. The E-RMF is considered to be one of the most successful model to reproduce the ground state properties of not only $\beta-$stable nuclei but also predicts quite reasonably the properties of drip-lines and superheavy nuclei \cite{kumar18,kumar17}. During last few decades, the application of this formalism to nuclear astrophysics is at forefront. It predicts the structure of NS and explains the tidal deformability satisfactorily \cite{malik2018}. The energy density functional for a nucleon-meson interacting system is given as \cite{kumar18}:
%%%%%%
\begin{widetext}
\begin{eqnarray}
{\cal E}({r})&=&\sum_{\alpha=p,n} \varphi_\alpha^\dagger({r})\Bigg\{-i \mbox{\boldmath$\alpha$} \!\cdot\!\mbox{\boldmath$\nabla$}+\beta \bigg[M-\Phi (r)-\tau_3 D(r)\bigg]+ W({r})+\frac{1}{2}\tau_3 R({r})+\frac{1+\tau_3}{2} A({r})-\frac{i\beta\mbox{\boldmath$\alpha$}}{2M}\!\cdot\!\bigg(f_\omega\mbox{\boldmath$\nabla$}W({r})
\nonumber\\
&&
+\frac{1}{2}f_\rho\tau_3 \mbox{\boldmath$\nabla$}R({r})\bigg)\Bigg\} \varphi_\alpha(r)+\left(\frac{1}{2}+\frac{\kappa_3}{3!}\frac{\Phi({r})}{M}+\frac{\kappa_4}{4!}\frac{\Phi^2({r})}{M^2}\right)\frac{m_s^2}{g_s^2}\Phi^2({r})-\frac{\zeta_0}{4!}\frac{1}{g_\omega^2 }W^4({r})+\frac{1}{2g_s^2}\left(1+\alpha_1\frac{\Phi({r})}{M}\right) \bigg(\mbox{\boldmath $\nabla$}\Phi({r})\bigg)^2
\nonumber\\
&&
-\frac{1}{2g_\omega^2}\left( 1 +\alpha_2\frac{\Phi({r})}{M}\right)\bigg(\mbox{\boldmath$\nabla$} W({r})\bigg)^2-\frac{1}{2}\left(1+\eta_1\frac{\Phi({r})}{M}+\frac{\eta_2}{2} \frac{\Phi^2({r})}{M^2}\right)\frac{m_\omega^2}{g_\omega^2} W^2({r})-\frac{1}{2e^2} \bigg( \mbox{\boldmath $\nabla$} A({r})\bigg)^2-\frac{1}{2g_\rho^2} \bigg( \mbox{\boldmath $\nabla$} R({r})\bigg)^2
\nonumber\\
&& 
-\frac{1}{2} \left( 1 + \eta_\rho \frac{\Phi({r})}{M}\right)\frac{m_\rho^2}{g_\rho^2} R^2({r}) -\Lambda_{\omega}\bigg(R^{2}(r)\times W^{2}(r)\bigg)+\frac{1}{2 g_{\delta}^{2}}\left( \mbox{\boldmath $\nabla$} D({r})\right)^2+\frac{1}{2}\frac{{m_{\delta}}^2}{g_{\delta}^{2}}D^{2}(r)\;,
\label{edf}
\end{eqnarray}
\end{widetext}
%%%%%%
Here, $\Phi$, $W$, $R$ and $D$ are the re-defined fields for $\sigma$, $\omega$, $\rho$ and $\delta$ mesons given as $\Phi = g_s\sigma_{0} $, $W = g_\omega \omega_{0}$, $R = g_\rho \vec{\rho_{0}}$ $^\mu$ and $D=g_{\delta}\delta_{0}$, respectively. $M$, $m_{\sigma}$, $m_{\omega}$, $m_{\rho}$ and $m_{\delta}$ are the masses of nucleon, $\sigma$, $\omega$, $\rho$ and $\delta$ mesons, respectively. From Eq. (\ref{edf}), we obtain the energy density ${\cal{E}}_{nucl.}$ \cite{kumar18,kumar17} by considering that the exchange of mesons create an uniform field, where the nucleon oscillates in a simple harmonic motion. From the E-RMF energy density, the equation of motions for the mesons and the nucleons are derived using the Euler-Lagrange equation. A set of coupled differential equations are obtained and solved self-consistently \cite{kumar18}. The scalar and vector densities,
\begin{eqnarray}
\rho_s(r)&=&\sum_\alpha \varphi_\alpha^\dagger({r})\beta\varphi_\alpha, \label{scaden}\\
\rho_v(r)&=&\sum_\alpha \varphi_\alpha^\dagger({r})\tau_{3}\varphi_\alpha\label{vecden},
\end{eqnarray}
are evaluated from the converged solutions within spherical harmonics. The vector density $\rho_v(r)$ is further used within CDFM to find out the weight function $|F(x)|^2$, which is an important quantity to calculate the incompressibility ($K^{A}$), symmetry energy ($S^{A}$), neutron pressure ($P^{A}$) and surface symmetry coefficient ($K_{sym}^{A}$) for the closed/semi-closed-shell even-even nuclei.\\
The expression for energy density of infinite and isotropic NM are obtained from the energy-momentum tensor:
\begin{eqnarray}
T_{\mu \nu} = &\sum_i \partial_\nu \phi_{i} \frac{\partial \cal L}{\partial (\partial^\mu \phi_{i})} -g_{\mu \nu} {\cal L}.
\label{tmunu}
\end{eqnarray}
The zeroth component of the energy-momentum tensor $T_{00}$ gives the energy density of the system as a function of baryon density as:
%%%%%%
\begin{eqnarray}
{\cal{E}}(k)_{nucl.}&=&\frac{2}{(2\pi)^{3}}\int d^{3}k E_{i}^\ast (k)+
\frac{ m_{s}^2\Phi^{2}}{g_{s}^2}\Bigg(\frac{1}{2}+\frac{\kappa_{3}}{3!}
\frac{\Phi }{M} + \frac{\kappa_4}{4!}\frac{\Phi^2}{M^2}\Bigg)
\nonumber\\
&&
+\rho_b W-\frac{1}{4!}\frac{\zeta_{0}W^{4}}
{g_{\omega}^2}
-\frac{1}{2}m_{\omega}^2\frac{W^{2}}{g_{\omega}^2}\Bigg(1+\eta_{1}\frac{\Phi}{M}+\frac{\eta_{2}}{2}\frac{\Phi ^2}{M^2}\Bigg)
\nonumber\\
&&
+\frac{1}{2}\rho_{3}R
-\frac{1}{2}\Bigg(1+\frac{\eta_{\rho}\Phi}{M}\Bigg)\frac{m_{\rho}^2}{g_{\rho}^2}R^{2}
-\Lambda_{\omega}  (R^{2}\times W^{2})
\nonumber\\
&&
+\frac{1}{2}\frac{m_{\delta}^2}{g_{\delta}^{2}}D^{2},
\label{enm}
\end{eqnarray}
%%%%%%%%%%%%%%%%%%%%%%%%%%%%%%%%%%

%%%%%%%%%%%%%%%%%%%%%%%%%%%%%%%%%%%%%%%%%%%%%%%%%%%%
\begin{figure}
\centering
\includegraphics[width=1\columnwidth]{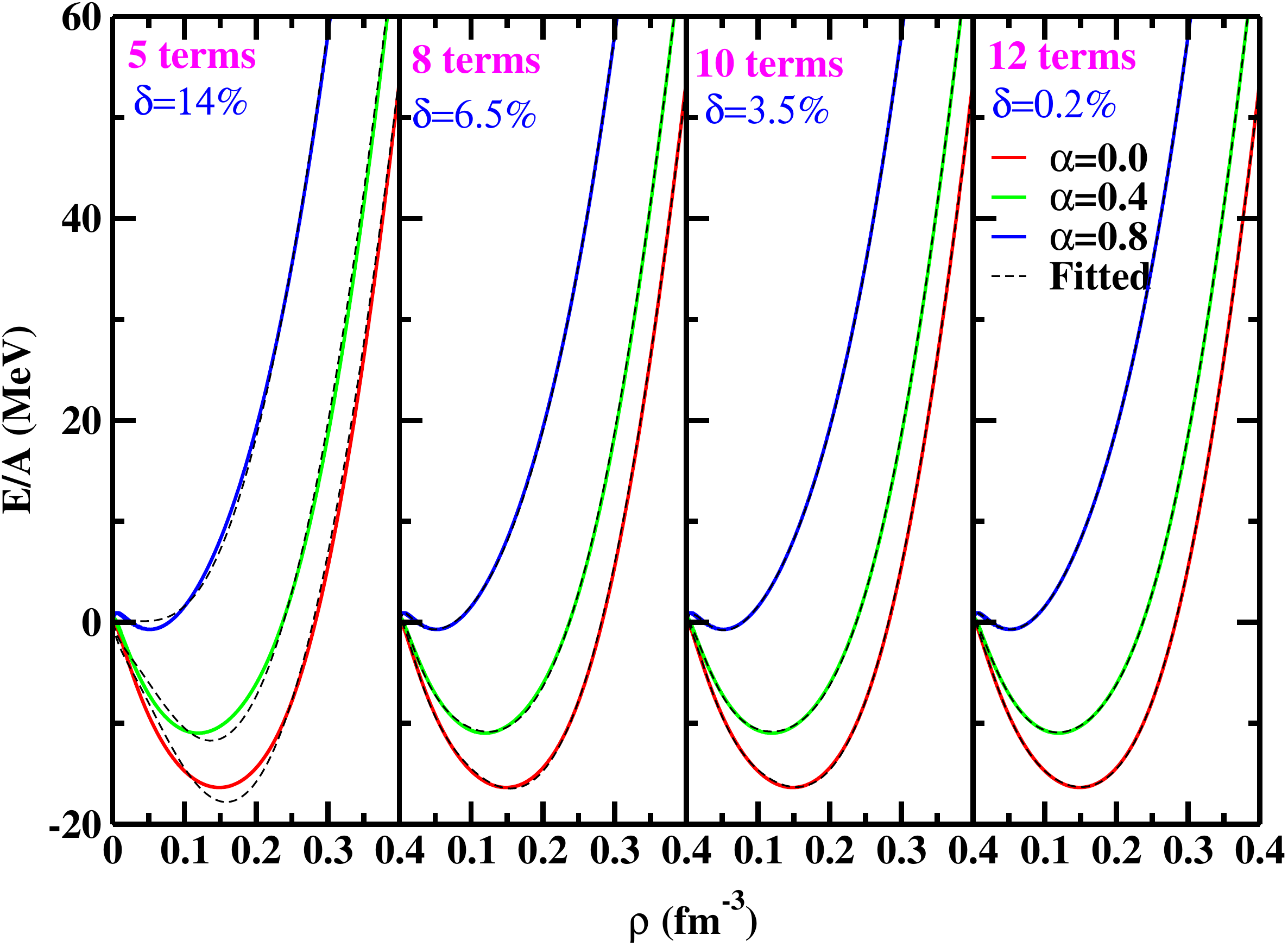}
\caption{(Color online) The numerical fitting of nuclear matter E/A as a function of baryon number density for NL3 parameter set using equation (\ref{efitting}) with a different number of terms. The $\delta$ represents the mean deviation between the RMF and fitted data. Here, 5, 8, and 12 terms in the fitting stand for the summation of $i$ from 3 to 7, 10, and 14 as given in Eq. (\ref {efitting}). More details can find in the text.}
\label{NMFitting}
\end{figure}
%%%%%%%%%%%%%%%%%%%
\begin{figure}
\centering
\includegraphics[width=1\columnwidth]{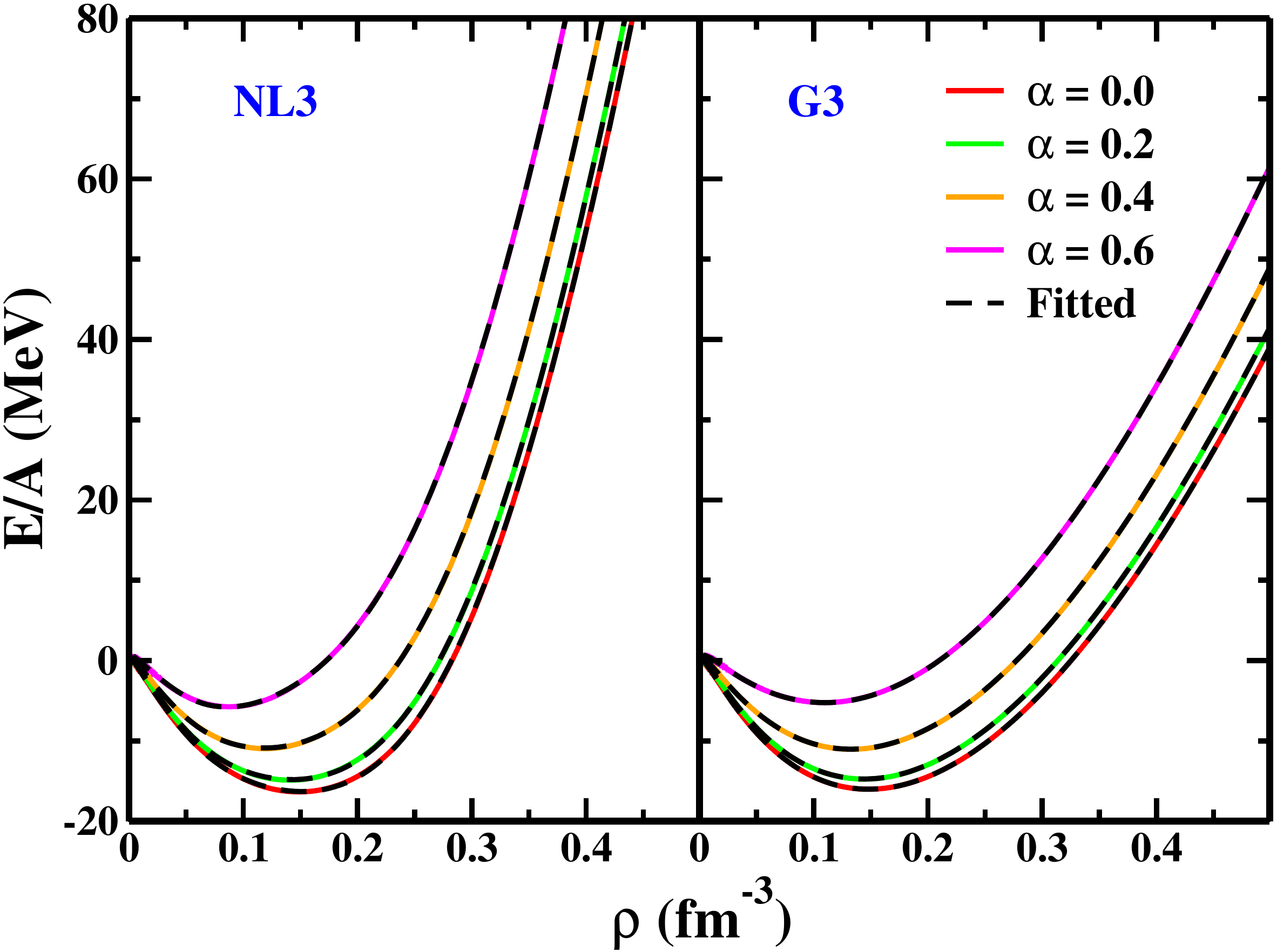}
\caption{(Color online) The nuclear matter E/A as a function of baryon number density for different asymmetry $\alpha = \frac{\rho_n-\rho_p}{\rho_n+\rho_p}$. For symmetric NM $\alpha=0$ and $\alpha=1$ for pure neutron matter.}
\label{NMEOS}
\end{figure}
%%%%%%%%%%%%%%%%%%%%%%%%%%%%%%%%%%%%%%%%%%%%%%%%%%%%%%%%

\subsection{Fitting procedure }\label{fitting}
The important part of the present calculation is to convert the NM quantities [Eq. (\ref{enm})] from momentum, space to coordinate space, i.e. the reconstruction of NM quantities at local density. The results of our calculations are shown in Fig. \ref{NMEOS} for NL3 and G3 parameter sets. The NL3 set gives a stiff equation of state (EOS) as compared to the G3 force. This is because the NM incompressibility $K$ at the saturation for NL3 is 271.76 MeV, and that of G3 is 243.96 MeV. We consider that the NM is composed of tiny spherical pieces described by a local density function $\rho_0(x) = 3A/4\pi x^3$. Using this consideration, the fitted binding energy function (Fig. \ref{NMEOS}) of E-RMF is embedded in the following equation,
%%%%%%%%%%%%%%%%%%%%%%%%%%
\begin{eqnarray}
{\cal E}(x) & = & C_k \rho_0^{2/3}(x) + \sum_{i=3}^{14} (b_i + a_i \alpha^2) \rho_0^{i/3}(x),
\label{efitting}
\end{eqnarray}
where $C_k$ is the kinetic energy coefficient given as $C_k = 37.53 [(1+\alpha)^{5/3} + (1-\alpha)^{5/3}]$ within the Thomas-Fermi approach. To find out the exact nature of the $E/A$ in position space, we use a polynomial fitting which consists of several numbers of terms (Eq. (\ref{efitting})). The fitted graph using the different number of terms for NL3 parameter set along with the mean deviation coefficient $\delta$ is depicted in Fig. \ref{NMFitting}. The mean deviation is calculated using the formula $\delta=\Big[\sum_{j=1}^{N} (E/A)_{j,Fitted} - (E/A)_{j,RMF}\Big]/N $. Here, $(E/A)_{j, Fitted}$ is the binding energy obtained from the polynomial fitting, and  $(E/A)_{j, RMF}$ is the binding energy per nucleon from the RMF functional with N being the number of data points. Firstly, we fitted our data using 5 terms in the expansion of Eq. (\ref{efitting}) (means $i$ runs from 3 to 7) and found that the deviation of fitted data is around 14 percent from the actual value. . Similarly, for 8, 10, and 12 terms, the deviation are 6.5\%, 3.5\%, and 0.2\%, respectively. We inspect that as we increase the number of terms in the expansion of Eq. (\ref{efitting}), the deviation of the fitting gets reduced considerably and the fitting converges to more candidness. We also observe that around 10-12 terms are required in the expansion for an appropriate fitting of most of the available RMF parameter sets. For the best fit, we take 12 number of terms in the present calculation, and the coefficients are obtained from the polynomial fitting, whose values are given in Table \ref{table1}.
%It is to be noted that one can restrict the number of coefficients %to a small number, but any arbitrary terms do not affect the model %as long as the fitting is perfect. \\
%%

The NM parameters $K^{NM}$, $S^{NM}$, $L_{sym}^{NM}$ and $K_{sym}^{NM}$ are obtained from the following standard relations \cite{kumar18,Chen2014}
%%%%%%
\begin{eqnarray}
K^{NM}&=&9\rho^2\frac{\partial^2}{\partial \rho^2} \bigg(\frac{\cal E}{\rho}\bigg)\Big|_{\rho=\rho_0} \label{knm},\\
S^{NM}&=&\frac{1}{2}\frac{\partial^2 ({\cal E}/\rho)}{\partial\alpha^2}\Big|_{\alpha=0},\label{snm}\\
L_{sym}^{NM}&=&3\rho\frac{\partial S(\rho)}{\partial\rho}\Big|_{\rho=\rho_0} = \frac{3P}{\rho_{0}},\label{lsymnm}\\
K_{sym}^{NM}&=&9\rho^2\frac{\partial^2 S(\rho)}{\partial\rho^2}\Big|_{\rho=\rho_0}.\label{ksymnm}
\end{eqnarray}
%%%%%%
which are given as follow using Eq. (6)
%%%%%%
\begin{eqnarray}
K^{NM} &=& -150.12\,\rho_0^{2/3}(x) + \sum_{i=4}^{14} i\, (i-3)\, b_i\, \rho_0^{i/3}(x), \label{eqA}\\
S^{NM} &=& 41.7\,\rho_0^{2/3}(x) + \sum_{i=3}^{14} a_i\, \rho_0^{i/3}(x), \label{eqB}\\
L_{sym}^{NM} &=& 83.4\,\rho_0^{2/3}(x) + \sum_{i=3}^{14} i\, a_i\, \rho_0^{i/3}(x), \label{eqC}\\
K_{sym}^{NM} &=& -83.4\,\rho_0^{2/3}(x) + \sum_{i=4}^{14} i\, (i-3)\, a_i\, \rho_0^{i/3}(x), \label{eqD}
\end{eqnarray}

%%%%%%%%%%%%%%%%%%%%%%%%%%%%%%%%%%%%%%%%%%%%%%%%%
\begin{table}
\caption{The coefficients of the analytical expression for NM
binding energy per particle as a function of density $\rho(x)$ and the asymmetric factor $\alpha=\frac{\rho_n-\rho_p}{\rho_n+\rho_p}$. The values are given for NL3 and G3 parameter sets.}
\renewcommand{\tabcolsep}{0.7cm}
\renewcommand{\arraystretch}{1.5}
\label{table1}
\begin{tabular}{ccccccc}
\hline \hline
& NL3  & G3  \\
\hline
b3 & -3449.92  & -490.15 \\ %\hline
b4 &  93386.65  & -465.80 \\ %\hline
b5 & -1233527.10  & 7107.17 \\ %\hline
b6 & 9041665.48  & -53960.91 \\ %\hline
b7 & -41166214.95  & 284155.27 \\ %\hline
b8 & 123164197.67  & -938303.73 \\ %\hline
b9 & -248225071.34 & 2066363.99 \\ %\hline
b10 & 338087637.81  & -3133853.65 \\ %\hline
b11 & -305682367.52 & 3246326.72 \\ %\hline
b12 & 174988863.34 & -2188895.63 \\ %\hline
b13 & -57095582.73 & 861872.30 \\ %\hline
b14 & 8030800.13 & -149719.19 \\ %\hline
a3 & -1098.99 & 391.87 \\ %\hline
a4 & 43110.63 & -5565.05 \\ %\hline
a5 & -636205.52 & 80413.45 \\ %\hline
a6 & 5283025.49  & -639847.05 \\ %\hline
a7 & -27638744.86 & 3139872.96 \\ %\hline
a8 & 96251325.61 & -10022304.90 \\ %\hline
a9 & -228719960.80 & 21277231.89 \\ %\hline
a10 & 372188746.60 & -30256280.52 \\ %\hline
a11 & -407653299.90 & 28503817.38 \\ %\hline
a12 & 286972591.80 & -17096240.48 \\ %\hline
a13 & -117150348.00 & 5921783.83 \\ %\hline
a14 & 21061682.62 & -903228.53 \\
\hline \hline
\end{tabular}
\label{tab1}
\end{table}
%%%%%%%%%%%%%%%%%%%%%%%%%%%%%%%%%%%%%%%%%%%%%%%%%%

The densities of closed/semi-closed-shell spherical nuclei $^{16}$O, $^{40,48}$Ca, $^{56}$Ni, $^{90}$Zr, $^{116}$Sn and $^{208}$Pb are calculated using E-RMF formalism. These densities are used as input in the CDFM (described in the following sub-section) to calculate the weight function, which is a key quantity acting as a bridge between NM parameters in $x-$space and finite nuclei in $r-$space (using LDA). To match with the $r-$ and $x-$space together, we construct the total density of the nucleus with the superposition of an infinite number of $Fluctons$, following the approach of CDFM discussed below.
%%%%%%%%%%
\subsection{Coherent Density Fluctuation model}\label{CDFM}
In the CDFM, we use the NM quantities $K^{NM}$, $S^{NM}$, $L_{sym}^{NM}$,
and $K_{sym}^{NM}$ from Eqs. (\ref{eqA}--\ref{eqD}) to extract their values for finite nuclei \cite{anto1,anto2,anto3,anto4}. Within CDFM, the one-body density matrix (OBDM) $\rho$ ({\bf r, r$'$}) of a finite nucleus is written as the coherent superposition of OBDM $\rho_x$ ({\bf r}, {\bf r$'$}) for spherical pieces of NM  termed as {\it Fluctons} \cite{bhu18,gad11},
\begin{equation}
\rho_x ({\bf r}) = \rho_0 (x)\, \Theta (x - \vert {\bf r} \vert),
\label{denx}
\end{equation}
with $\rho_o (x) = \frac{3A}{4 \pi x^3}$. The generator coordinate x is the radius of a sphere consisting of Fermi gas having all the  A nucleons distributed uniformly within it. It is suitable to apply for such a system the OBDM expressed as below \cite{bhu18,anto2,gad11,gad12},
\begin{equation}
\rho ({\bf r}, {\bf r'}) = \int_0^{\infty} dx \vert F(x) \vert^2 \rho_x
({\bf r}, {\bf r'}),
\label{denr}
\end{equation}
where, $\vert F(x) \vert^2 $ is the weight function (WF). The coherent superposition of OBDM $\rho_x ({\bf r}, {\bf r'})$ is given as:
\begin{eqnarray}
\rho_x ({\bf r}, {\bf r'}) &=& 3 \rho_0 (x) \frac{J_1 \left( k_f (x) \vert
{\bf r} - {\bf r'} \vert \right)}{\left( k_f (x) \vert {\bf r} - {\bf r'}
\vert \right)} \nonumber \\
&&\times \Theta \left(x-\frac{ \vert {\bf r} + {\bf r'} \vert }{2} \right),
\label{denrr}
\end{eqnarray}
where J$_1$ is the first order spherical Bessel function and $k_{f}$is the Fermi momentum of nucleons inside the $Flucton$ having
radius $x$ and  $k_f (x)=(3\pi^2/2\rho_0(x))^{1/3} =\gamma/x$, ($\gamma\approx
1.52A^{1/3}$). The Wigner distribution function of the OBDM of Eq. (\ref{denrr}) is given by,
\begin{eqnarray}
W ({\bf r}, {\bf k}) =  \int_0^{\infty} dx\, \vert F(x) \vert^2\, W_x ({\bf r}, {\bf k}).
\label{wing}
\end{eqnarray}
Here, $W_x ({\bf r}, {\bf k})=\frac{4}{8\pi^3}\Theta (x-\vert {\bf r} \vert)\Theta (k_F(x)-\vert {\bf k} \vert)$.
The density $\rho$ (r) in terms of the WF within the CDFM approach is:
\begin{eqnarray}
\rho (r) &=& \int d{\bf k} W ({\bf r}, {\bf k}) \nonumber \\
&& = \int_0^{\infty} dx\, \vert F(x) \vert^2\, \frac{3A}{4\pi x^3} \Theta(x-\vert{\bf r} \vert),
\label{rhor}
\end{eqnarray}
%%%%%%
which is normalized to A, i.e.,  $\int \rho ({\bf r})d{\bf r} = A$. In the $\delta$-function limit, the Hill-Wheeler integral equation, that is the differential equation for the WF in the generator coordinate is obtained \cite{anto1}. The $|F(x)|^2$ for a given density $\rho$ (r) is expressed as
%%%%%%
\begin{equation}
|F(x)|^2 = - \left (\frac{1}{\rho_0 (x)} \frac{d\rho (r)}{dr}\right)_{r=x},
\label{wfn}
\end{equation}
%%%%%%
with $\int_0^{\infty} dx \vert F(x) \vert^2 =1$. We refer \cite{bhu18,anto1,anto2,gad11,gad12} for a detailed analytical derivation. The CDFM allows us to make a transition from the properties of NM to those of finite nuclei. The finite nuclear incompressibility $K^{A}$, symmetry energy $S^{A}$, neutron pressure $P^{A}$ and surface incompressibility $K_{sym}^{A}$ for a finite nucleus are calculated by weighting the corresponding quantities for infinite NM within the CDFM, as given below \cite{anto4,gad11,gad12,fuch95,anto17}
%%%%%%
\begin{eqnarray}
K^{A}=  \int_0^{\infty} dx\, \vert F(x) \vert^2 \ K^{NM} (\rho (x)).
\label{K0}
\end{eqnarray}
\begin{eqnarray}
P^{A} =  \int_0^{\infty} dx\, \vert F(x) \vert^2\, P^{NM} (\rho (x)),
\label{p0}
\end{eqnarray}
\begin{eqnarray}
S^{A}= \int_0^{\infty} dx\, \vert F(x) \vert^2\, S^{NM} (\rho (x)) ,
\label{s0}
\end{eqnarray}
\begin{eqnarray}
L_{sym}^{A}= \int_0^{\infty} dx\, \vert F(x) \vert^2 \,L_{sym}^{NM} (\rho (x)) ,
\label{L0}
\end{eqnarray}
\begin{eqnarray}
K_{sym}^{A} =  \int_0^{\infty} dx\, \vert F(x) \vert^2 \ K_{sym}^{NM} (\rho (x)),
\label{k0}
\end{eqnarray}
%%%%%%
The $K^{A}$, $P^{A}$, $S^{A}$, $L_{sym}^{A}$, and $K_{sym}^{A}$ in  Eqs. ($\ref{K0}-$\ref{k0}) are the surface weighted average of the corresponding NM quantities in the LDA limit for finite nuclei.
%%%%%%%%%%%%%%%%%%%%%%%%%%%%%%%%%%%%%%%%%%%%%%%%%%%%
\section{Results and discussion}
\label{results}
In the present work, we have derived the relativistic density functional expression from the NM calculations for different neutron-proton asymmetry ($\alpha$) using the NL3 and the recently developed G3 parameter sets. It is quite important to note that the Br$\ddot{u}$ckner density functional \cite{bruk68} used within a CDFM to calculate symmetry energy and related observables in previous studies \cite{gad16,bhu18,abdul19} is inadequate. The underlying reason is that the NM saturation curves fitted by Br$\ddot{u}$ckner energy density functional could not reproduce the empirical values of saturation density ($\rho \sim 0.15$ fm$^{-3}$) and binding energy per nucleon (E $\sim$ 16 MeV) of NM (see Fig.1 of Ref. \cite{bruk68}). In other words, Br$\ddot{u}$ckner energy density functional could not sort out the Coster-Band problem \cite{coester}. This issue was rectified partially by the inclusion of three-body force in the nucleon-nucleon potential \cite{fujita, Pandhere}. To address this issue, we emphasize to obtain relativistic energy density functional since the RMF Lagrangian with non-linear terms mimics the three-body effect in the nuclear potential and resolves the Coster-Band problem.

%%%%%%%%%%%%%%%%%%%
\begin{figure}
\centering
\includegraphics[width=0.9\columnwidth]{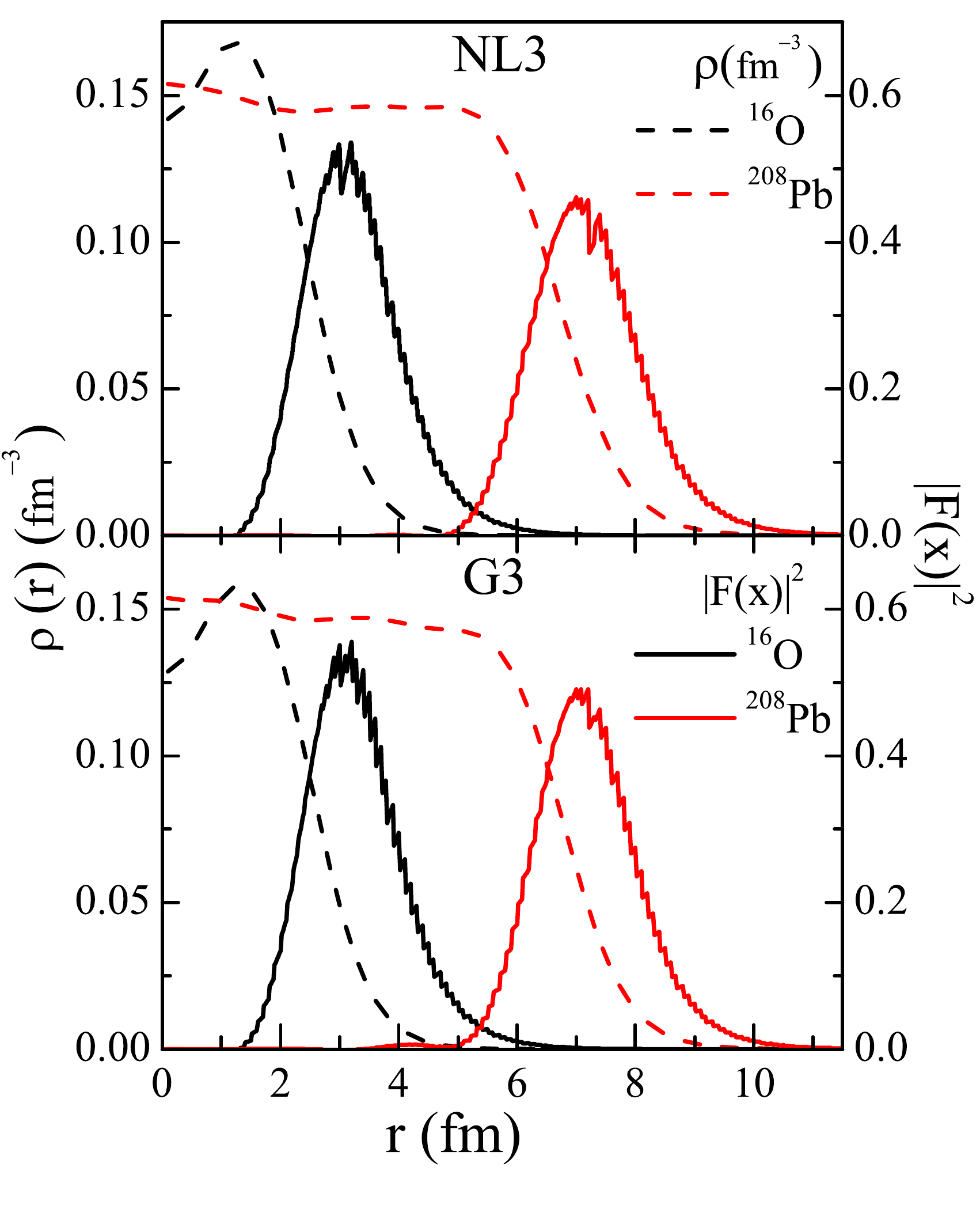}
\vspace{0.1 cm}
\caption{(color online) The densities (dotted lines) and weight functions (solid lines) of $^{16}$O and $^{208}$Pb for NL3 (upper panel) and G3 (lower panel) parameter sets.}
\label{BE}
\end{figure}
%%%%%%%%%%%%%%%%%%%%%%%
\begin{table}
\caption{The properties such as nuclear incompressibility $K^{A}$, symmetric energy $S^{A}$, neutron pressure $P^{A}$ = $\rho_{0}$ $L_{sym}^{A}$/3, slope $L_{sym}^{A}$ and curvature $K_{sym}^{A}$ of nuclei, with RMF density functional, for NL3 and G3 parameter sets. The dimension of all the parameters are in MeV.}
\renewcommand{\tabcolsep}{0.07cm}
\renewcommand{\arraystretch}{1.5}
\label{table2}
\begin{tabular}{ccccccccccc}
\hline \hline
NL3& $^{16}$O & $^{40}$Ca & $^{48}$Ca & $^{56}$Ni & $^{90}$Zr & $^{116}$Sn  & $^{208}$Pb  \\
\hline
$K^{A}$ & 618.05  & 584.11  & 564.86  & 627.33  & 450.87   & 476.46  & 411.03  \\ 
$P^{A}$ & 9.70    & 8.30    & 7.84    & 8.68    & 6.90     & 7.24    & 6.68  \\
$S^{A}$ & 40.30   & 39.40   & 38.83   & 41.66   & 36.97    & 38.38   & 37.28  \\ 
$L_{sym}^{A}$   & 118.89  & 119.79  & 120.54  & 130.30  & 117.01  & 121.20  & 117.95  \\ 
$K_{sym}^{A}$   & 32.42   & -10.92  & -3.18   & -4.37   & 33.30   & 31.34   & 47.99  \\
\hline
G3 & $^{16}$O & $^{40}$Ca & $^{48}$Ca & $^{56}$Ni & $^{90}$Zr & $^{116}$Sn  & $^{208}$Pb  \\
\hline
$K^{A}$ & 258.87  & 262.09  & 270.26  & 279.33  & 253.87  & 249.74  & 238.85  \\ 
$P^{A}$ & 3.39    & 3.07    & 3.08    & 3.12    & 2.75    & 2.68    & 2.52  \\ 
$S^{A}$ & 30.12   & 30.43   & 31.15   & 31.98   & 30.88   & 30.87   & 30.53  \\ 
$L_{sym}^{A}$   & 51.98   & 51.92   & 52.71   & 53.57   & 51.40   & 51.18   & 50.28  \\ 
$K_{sym}^{A}$   &-103.23  &-89.43   &-89.52   &-90.00   &-91.55   &-93.07   &-96.39  \\ \hline \hline
\end{tabular}
\end{table}
%%%%%%%%%%%%
\begin{table}
\caption{The properties such as symmetric energy $S^{A}$, neutron pressure $P^{A}$ and curvature $K_{sym}^{A}$  of nuclei for NL3 and G3 parameter sets using Br$\ddot{u}$ckner's density functional within CDFM. The dimension of all the parameters are in MeV.}
\renewcommand{\tabcolsep}{0.05 cm}
\renewcommand{\arraystretch}{1.5}
\label{table3}
\begin{tabular}{ccccccccccc}
\hline \hline
NL3& $^{16}$O & $^{40}$Ca & $^{48}$Ca & $^{56}$Ni & $^{90}$Zr & $^{116}$Sn  & $^{208}$Pb  \\
\hline
$S^{A}$ & 25.33 &	27.28	& 28.36  & 29.91	& 29.13  & 29.38 & 29.28\\ 
$P^{A}$ & -0.66 &	0.05	&	-0.05	&	0.02	&	0.95	&	1.25	&	1.57  \\ 
$K_{sym}^{A}$ & -281.34	& -275.28	& -296.82	& -324.29	& -267.48	& -257.88	& -236.89  \\
\hline
G3 & $^{16}$O & $^{40}$Ca & $^{48}$Ca & $^{56}$Ni & $^{90}$Zr & $^{116}$Sn  & $^{208}$Pb  \\
\hline
$S^{A}$ & 25.39	& 27.28	    & 28.42	& 29.40	& 29.05	& 29.19	& 29.12\\ 
$P^{A}$           & -0.41	& 0.31	& 0.32	& 0.53	& 1.23	& 1.48	& 1.61  \\ 
$K_{sym}^{A}$ & -262.45	& -261.99	& -227.39	& -288.89	& -251.25	& -241.56	& -229.63 \\
\hline
\end{tabular}
\end{table}
%%%%%%%%%%%%
The NM binding energy per particle for variable $\alpha$ within the RMF models (NL3 and G3 parameter sets) are shown in Fig. \ref{NMEOS}. These curves for different values of $\alpha$ are fitted with a polynomial [Eq. (\ref{efitting})] where the first term is the kinetic energy taken from Thomas-Fermi approximation and the second term presents the potential energy.  In the potential energy terms, the obtained coefficients $a_{i}$ and $b_{i}$ are shown in Table \ref{table1}. It is important to note that the expression $E/A$ of NM in Eq. (\ref{enm}) is obtained as a function of the density of spherical pieces of NM $\rho_0(x)$, called {\it Fluctons}, as the expansion variable. The Eq. (\ref{efitting}) is, in fact, equivalent energy expression of Eq. (\ref{enm}) and therefore the subsequent expressions of symmetry energy and other quantities (Eqs. (\ref{K0}--\ref{k0})) obtained from Eq. (\ref{efitting}) also encompass the relativistic characteristics. %%%%%%%%%
\begin{figure}
\centering
\includegraphics[width=0.9\columnwidth]{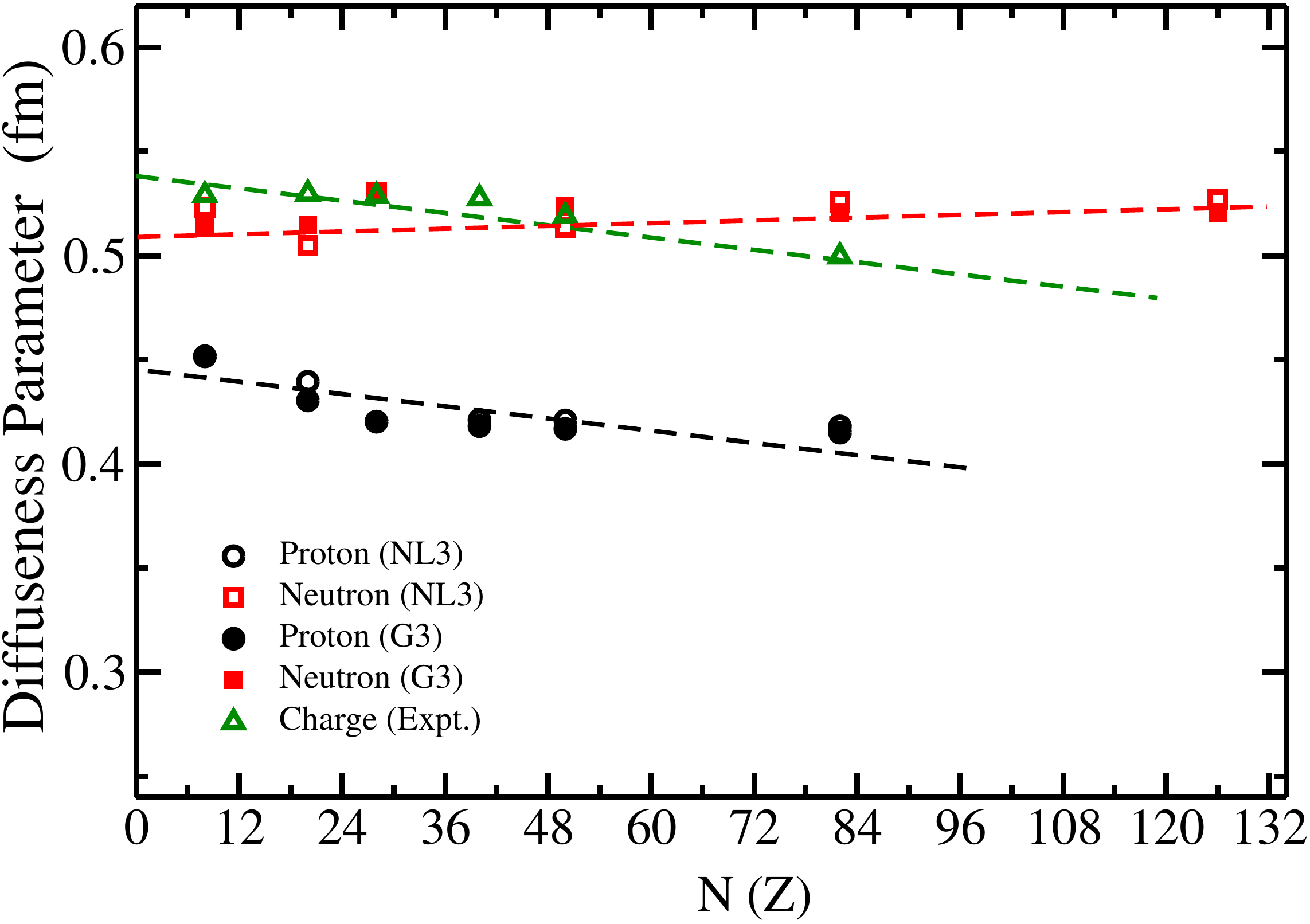}
\caption{(color online) The proton (circle) and neutron (square) surface diffusion parameter for the considered nuclei in Table \ref{table2} for NL3 (open symbol) and G3 (solid symbol) parameter sets along with the experimental charge (open triangle) diffuseness estimates \cite{varies87,nadj94}. See the text for more details.}
\label{diffuse}
\end{figure}
%%%%%%%%%%%%%
Further, within CDFM, NM  parameters together with weight functions will be used to evaluate the corresponding quantities of finite closed/semi-closed-shell $^{16}$O, $^{40,48}$Ca, $^{56}$Ni, $^{90}$Zr, $^{116}$Sn and $^{208}$Pb nuclei (see Eqs. (\ref{K0}--\ref{k0})). The self-consistently calculated RMF densities are the crucial ingredient to find weight function of nuclei and both the density and weight function for $^{16}$O and $^{208}$Pb nuclei are shown in Fig. \ref{BE} as representative cases. The dashed lines present the density of nuclei, and the solid lines depict the variation of weight function. It is evident that although the density at the centre of $^{16}$O and $^{208}$Pb nuclei is high, the value of weight function in that region is very small. Moreover, the weight function exhibits a bell shape with a maxima, which corresponds to the surface region where the density is significantly reduced compared to the central region. Hence, in the surface region, the weight function makes a substantial contribution to the calculation of symmetry energy and other quantities. Due to this reason, the symmetry energy, neutron pressure, etc. are labelled as surface properties. It is also important to note that the total symmetry energy $S$ can be made a partition into volume $S_V$ and surface $S_S$ components with an analogy of two connected capacitors. A detailed discussion can be obtained in Refs. \cite{kaurnpa,dani03,dani04,dani06,diep07,dani09,anto18}.\\
%%%%%%%%%%%%%%%%%%%%%%
\begin{figure*}
\centering
\includegraphics[width=1.8\columnwidth]{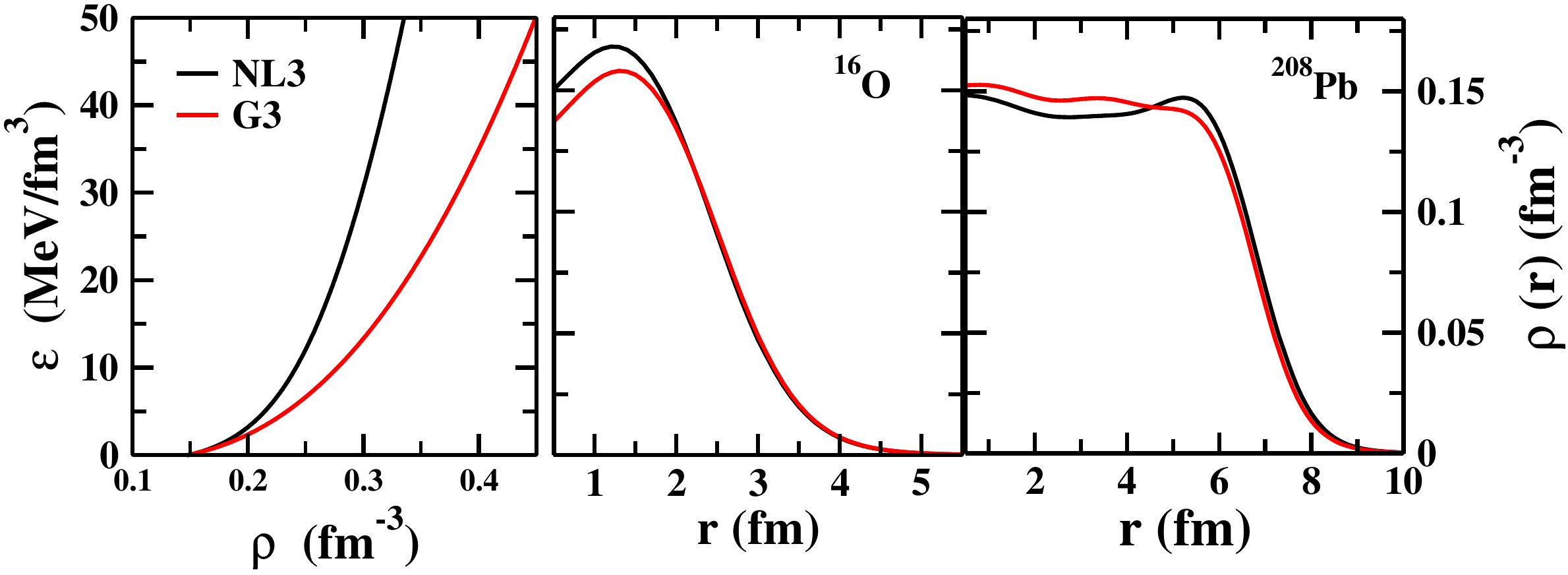}
\vspace{0.1 cm}
\caption{(color online) Left panel: The energy density ($\cal{E}$) vs density ($\rho$). Right panel: The  density distribution of $^{16}$O and $^{208}$Pb for NL3 and G3 parameter sets.}
\label{eosden}
\end{figure*}
%%%%%%%%%%%%%%%%%%%%%%%

\noindent
To determine the surface effect on the symmetry energy of finite nuclei, we calculate the surface diffuseness parameters for all the nuclei by using the nucleon density distributions. The equivalent nuclear surface diffuseness parameter can be obtained using the relation $a_i \approx -\rho_i/ \frac{d\rho_i}{dr}$, where i stands for proton-($a_p$), neutron-($a_n$) and charge-density ($a_{ch}$) density distributions \cite{bhu20} and reference therein. The results for the surface diffuseness parameter are obtained for proton (circle) and neutron (square) by using the respective densities from the RMF approach for NL3 (open symbol) and G3 (solid symbol) parameter sets. The experimental diffuseness parameter (open triangle) are estimated from the charge density distribution \cite{varies87,nadj94}. All the results along with the experimental estimates are given in Fig. \ref{diffuse}. From the figure, one can observe that there is a constant difference in the calculated diffuseness parameter for proton and the experimental data for the charge. This is due to the charge effect in the proton density in the experiment, whereas we use proton number density in which the charge effect is absent. Comparing  the Fig. \ref{diffuse} and Table \ref{table2}, one can find a linear correlation among the neutron diffuseness parameter and symmetry energy. In other words, the trend appearing for the symmetry energy exactly reflects in the neutron surface diffuseness parameter. In contrast, the diffuseness parameter for proton follows the linear fall, which is consistent with the experimental data. Hence, the similar trends in the symmetry energy and surface diffuseness parameter for neutron show the implication of surface effects in terms of weight function for the study of NM quantities for finite nuclei.

The surface properties such as symmetry energy, neutron pressure, slope and curvature of symmetry energy, incompressibility of NM Eqs. (\ref{eqA}--\ref{eqD}) are computed using relativistic density functional Eqs. (\ref{efitting}). These are further folded with weight function to calculate the corresponding quantities of closed/semi-closed-shell even-even nuclei, which are shown in Table \ref{table2}. From the table, one can notice a wide range of finite nuclear incompressibility $K^{A}$ = 618.05 to 411.03 MeV and $K^{A}$ = 258.87 to 238.85 MeV for $^{16}$O - $^{208}$Pb with  NL3 and G3 parameter set, respectively. To understand this nature, we have given the density distribution of nucleons for both NL3 and G3 in Fig. \ref{eosden} for $^{16}$O and $^{208}$Pb as two representative cases. The energy density functional ($\cal{E}$) for both NL3 and G3 sets are also displayed in the left panel of the same figure. It is very much clear that the densities are significantly different for both the forces, mostly in the central region. In the case of G3, a comparatively flat distribution of nucleons appears as compared to NL3, predicting a small variation in the weight function for both the sets. However, if one analyzes the variation in the energy density $\cal{E}$ (or E/A) with density $\rho$ for both NL3 and G3 sets, a huge difference between these two sets are visible. For the NL3 case, the variation in energy density $\cal{E}$ with $\rho$ is substantially larger than G3. As a result, we get a small $K^A$ for G3 than NL3, although the NM incompressibility at the saturation is almost comparable for the two force parameters. 
From Eq. (\ref{knm}), we know that the incompressibility is the second derivative of EOS (E/A) for the density. Therefore, the incompressibility predicted by NL3 is more than the G3 for all nuclei. The variation in other quantities can also be explained taking into the behavior of E/A for various force parameters. From these results, we cannot conclude about the mass dependence of the finite nuclear incompressibility, for example, $K^{A}=$ 618, 627, and 411 MeV for $^{16}$O, $^{56}$Ni and $^{208}$Pb, respectively for NL3 case. Similar uncertainty in $K^{A}$ for different nuclei is also seen in the G3 case. However, for $P^{A}$, $S^{A}$ and $L_{sym}^{A}$, the variation is in a narrow range from $^{16}$O to $^{208}$Pb, i.e., minimum $P^{A}$ is 2.52 MeV for $^{208}$Pb and maximum is 3.39 MeV for $^{16}$O with G3. Furthermore, $K_{sym}^{A}$ varies a lot depending on both the force parameter as well as the mass of the nucleus. The very different values of $K^{A}$ and $K_{sym}^{A}$ for light and heavy nuclei indicate the structural dependence of the finite nuclei since the density distribution varies from the nucleus to nucleus.
%%% 

For the sake of comparison, the values of symmetry energy and other quantities of nuclei obtained using Br$\ddot{u}$ckner density functional within CDFM are shown in Table \ref{table3} for NL3 and G3 densities. The values of symmetry energy $S^{A}$ linearly increases with the mass number and the lies in the range 25-29 MeV.  Similar conclusions can be drawn for $P^{A}$ and $K_{sym}^{A}$ for these considered nuclei. We get $P^{A}$ is negative for $^{16}$O for both parameter sets, which can be correlated with the charge effect of proton for  N = Z nucleus. We also get negative $P^{A}$ for $^{48}$Ca for the NL3 case, which may be connected with the reduced charge radius of $^{48}$Ca in comparison to $^{40}$Ca, mainly an unsolved issue in the theoretical models. In comparison, the results using relativistic density functional along with mean-field densities for both the parameter sets give positive $P^{A}$ value for all the nuclei (see Table \ref{table2}). We get a large difference and/or a wide range of values for symmetry energy curvature $K_{sym}^{A}$ in Table \ref{table3}.  This difference is due to the small difference in the resultant energy density for relativistic and Br$\ddot{u}$kner energy density functional (see Table \ref{table2}). 
%%%%%%%%%%%%%%%%%%%%%%%%%%%%%%%%%%%%%%%%%%%%%%%%%%%%%%%%%%%%%%%%%%%%%%%%%%
\section{Summary and Conclusion}
\label{summary}
In brief, we have fitted the NM saturation curves for different values of asymmetry parameter, employing the E-RMF density functional with two well-known G3 and NL3 parameter sets, in the wake of the solution of the Coster-Band problem by relativistic approach. The newly fitted expression of E/A is used to find NM parameters such as incompressibility, neutron pressure, symmetry energy. These NM parameters, together with E-RMF densities, are used in the calculations of surface properties of magic nuclei ranging from light to heavy mass region, within the coherent density fluctuation model. The values of symmetry energy, neutron pressure, L-coefficient show small variation for different mass nuclei.
On the other hand, the incompressibility and surface incompressibility show large variation while moving from light $^{16}$O to heavy $^{208}$Pb nuclei for NL3 compared to the G3 case. This is due to the variation of E/A with density or EOS is stiff for NL3 than in the G3 case and as a result, the incompressibility (second derivative of E/A w.r.t. density) predicted by NL3 is more than G3 for all nuclei. Besides, the correlation between the diffuseness parameter and symmetry energy is discussed in the context of surface effects. The results using relativistic energy density functions are also compared with ones obtained using Br$\ddot{u}$kner energy density functional.

In other words, we discuss here that the direct use of Br$\ddot{u}$ckner energy functional to evaluate the effective nuclear surface properties is not adequate in the context of the Coster-Band issue. The present work provides the way to estimate the properties of $\beta$-stable and $\beta$-unstable nuclei more accurately by employing relativistic density functional within CDFM. These quantities are quite significant to probe the structural aspects of finite nuclei and can also facilitate to constrain the EOS of NM. Further, we also compare the RMF results with Br$\ddot{u}$ckner density functional, which reflects the model dependant on the results. A more systematic study using the present functional for the nuclei throughout the nuclear chart in this direction is highly welcomed. \\

\noindent 
{\bf Acknowledgement:} One of the authors (MB) is thankful to the Institute of Physics, Bhubaneswar for the local hospitality during the scientific visit. MB acknowledges the support from FAPESP Project No. 2017/05660-0, FOSTECT Project No. FOSTECT.2019B.04, and the CNPq - Brasil. \\

\bibliography{cdfm}

\begin{thebibliography}{42}%
\makeatletter
\providecommand \@ifxundefined [1]{%
 \@ifx{#1\undefined}
}%
\providecommand \@ifnum [1]{%
 \ifnum #1\expandafter \@firstoftwo
 \else \expandafter \@secondoftwo
 \fi
}%
\providecommand \@ifx [1]{%
 \ifx #1\expandafter \@firstoftwo
 \else \expandafter \@secondoftwo
 \fi
}%
\providecommand \natexlab [1]{#1}%
\providecommand \enquote  [1]{``#1''}%
\providecommand \bibnamefont  [1]{#1}%
\providecommand \bibfnamefont [1]{#1}%
\providecommand \citenamefont [1]{#1}%
\providecommand \href@noop [0]{\@secondoftwo}%
\providecommand \href [0]{\begingroup \@sanitize@url \@href}%
\providecommand \@href[1]{\@@startlink{#1}\@@href}%
\providecommand \@@href[1]{\endgroup#1\@@endlink}%
\providecommand \@sanitize@url [0]{\catcode `\\12\catcode `\$12\catcode
  `\&12\catcode `\#12\catcode `\^12\catcode `\_12\catcode `\%12\relax}%
\providecommand \@@startlink[1]{}%
\providecommand \@@endlink[0]{}%
\providecommand \url  [0]{\begingroup\@sanitize@url \@url }%
\providecommand \@url [1]{\endgroup\@href {#1}{\urlprefix }}%
\providecommand \urlprefix  [0]{URL }%
\providecommand \Eprint [0]{\href }%
\providecommand \doibase [0]{http://dx.doi.org/}%
\providecommand \selectlanguage [0]{\@gobble}%
\providecommand \bibinfo  [0]{\@secondoftwo}%
\providecommand \bibfield  [0]{\@secondoftwo}%
\providecommand \translation [1]{[#1]}%
\providecommand \BibitemOpen [0]{}%
\providecommand \bibitemStop [0]{}%
\providecommand \bibitemNoStop [0]{.\EOS\space}%
\providecommand \EOS [0]{\spacefactor3000\relax}%
\providecommand \BibitemShut  [1]{\csname bibitem#1\endcsname}%
\let\auto@bib@innerbib\@empty
%</preamble>
\bibitem [{\citenamefont {Lattimer}\ and\ \citenamefont
  {Prakash}(2000)}]{latimer2000}%
  \BibitemOpen
  \bibfield  {author} {\bibinfo {author} {\bibfnamefont {J.~M.}\ \bibnamefont
  {Lattimer}}\ and\ \bibinfo {author} {\bibfnamefont {M.}~\bibnamefont
  {Prakash}},\ }\href {\doibase https://doi.org/10.1016/S0370-1573(00)00019-3}
  {\bibfield  {journal} {\bibinfo  {journal} {Physics Reports}\ }\textbf
  {\bibinfo {volume} {333-334}},\ \bibinfo {pages} {121 } (\bibinfo {year}
  {2000})}\BibitemShut {NoStop}%
\bibitem [{\citenamefont {Lattimer}\ and\ \citenamefont
  {Prakash}(2001)}]{latimer01}%
  \BibitemOpen
  \bibfield  {author} {\bibinfo {author} {\bibfnamefont {J.~M.}\ \bibnamefont
  {Lattimer}}\ and\ \bibinfo {author} {\bibfnamefont {M.}~\bibnamefont
  {Prakash}},\ }\href {\doibase 10.1086/319702} {\bibfield  {journal} {\bibinfo
   {journal} {The Astrophysical Journal}\ }\textbf {\bibinfo {volume} {550}},\
  \bibinfo {pages} {426} (\bibinfo {year} {2001})}\BibitemShut {NoStop}%
\bibitem [{\citenamefont {Steiner}\ \emph {et~al.}(2005)\citenamefont
  {Steiner}, \citenamefont {Prakash}, \citenamefont {Lattimer},\ and\
  \citenamefont {Ellis}}]{steiner05}%
  \BibitemOpen
  \bibfield  {author} {\bibinfo {author} {\bibfnamefont {A.}~\bibnamefont
  {Steiner}}, \bibinfo {author} {\bibfnamefont {M.}~\bibnamefont {Prakash}},
  \bibinfo {author} {\bibfnamefont {J.}~\bibnamefont {Lattimer}}, \ and\
  \bibinfo {author} {\bibfnamefont {P.}~\bibnamefont {Ellis}},\ }\href
  {\doibase https://doi.org/10.1016/j.physrep.2005.02.004} {\bibfield
  {journal} {\bibinfo  {journal} {Physics Reports}\ }\textbf {\bibinfo {volume}
  {411}},\ \bibinfo {pages} {325 } (\bibinfo {year} {2005})}\BibitemShut
  {NoStop}%
\bibitem [{\citenamefont {Li}\ \emph {et~al.}(2008)\citenamefont {Li},
  \citenamefont {Chen},\ and\ \citenamefont {Ko}}]{li08}%
  \BibitemOpen
  \bibfield  {author} {\bibinfo {author} {\bibfnamefont {B.-A.}\ \bibnamefont
  {Li}}, \bibinfo {author} {\bibfnamefont {L.-W.}\ \bibnamefont {Chen}}, \ and\
  \bibinfo {author} {\bibfnamefont {C.~M.}\ \bibnamefont {Ko}},\ }\href
  {\doibase https://doi.org/10.1016/j.physrep.2008.04.005} {\bibfield
  {journal} {\bibinfo  {journal} {Physics Reports}\ }\textbf {\bibinfo {volume}
  {464}},\ \bibinfo {pages} {113 } (\bibinfo {year} {2008})}\BibitemShut
  {NoStop}%
\bibitem [{\citenamefont {Lynch}\ \emph {et~al.}(2009)\citenamefont {Lynch},
  \citenamefont {Tsang}, \citenamefont {Zhang} \emph {et~al.}}]{lynch09}%
  \BibitemOpen
  \bibfield  {author} {\bibinfo {author} {\bibfnamefont {W.}~\bibnamefont
  {Lynch}}, \bibinfo {author} {\bibfnamefont {M.}~\bibnamefont {Tsang}},
  \bibinfo {author} {\bibfnamefont {Y.}~\bibnamefont {Zhang}},  \emph
  {et~al.},\ }\href {\doibase https://doi.org/10.1016/j.ppnp.2009.01.001}
  {\bibfield  {journal} {\bibinfo  {journal} {Progress in Particle and Nuclear
  Physics}\ }\textbf {\bibinfo {volume} {62}},\ \bibinfo {pages} {427 }
  (\bibinfo {year} {2009})}\BibitemShut {NoStop}%
\bibitem [{\citenamefont {Tsang}\ \emph {et~al.}(2012)\citenamefont {Tsang},
  \citenamefont {Stone}, \citenamefont {Camera} \emph {et~al.}}]{tsang12}%
  \BibitemOpen
  \bibfield  {author} {\bibinfo {author} {\bibfnamefont {M.~B.}\ \bibnamefont
  {Tsang}}, \bibinfo {author} {\bibfnamefont {J.~R.}\ \bibnamefont {Stone}},
  \bibinfo {author} {\bibfnamefont {F.}~\bibnamefont {Camera}},  \emph
  {et~al.},\ }\href {\doibase 10.1103/PhysRevC.86.015803} {\bibfield  {journal}
  {\bibinfo  {journal} {Phys. Rev. C}\ }\textbf {\bibinfo {volume} {86}},\
  \bibinfo {pages} {015803} (\bibinfo {year} {2012})}\BibitemShut {NoStop}%
\bibitem [{\citenamefont {Horowitz}\ \emph {et~al.}(2014)\citenamefont
  {Horowitz}, \citenamefont {Brown}, \citenamefont {Kim} \emph
  {et~al.}}]{horo14}%
  \BibitemOpen
  \bibfield  {author} {\bibinfo {author} {\bibfnamefont {C.~J.}\ \bibnamefont
  {Horowitz}}, \bibinfo {author} {\bibfnamefont {E.~F.}\ \bibnamefont {Brown}},
  \bibinfo {author} {\bibfnamefont {Y.}~\bibnamefont {Kim}},  \emph {et~al.},\
  }\href {\doibase 10.1088/0954-3899/41/9/093001} {\bibfield  {journal}
  {\bibinfo  {journal} {Journal of Physics G: Nuclear and Particle Physics}\
  }\textbf {\bibinfo {volume} {41}},\ \bibinfo {pages} {093001} (\bibinfo
  {year} {2014})}\BibitemShut {NoStop}%
\bibitem [{\citenamefont {Antonov}\ \emph {et~al.}(2016)\citenamefont
  {Antonov}, \citenamefont {Gaidarov}, \citenamefont {Sarriguren},\ and\
  \citenamefont {Moya~de Guerra}}]{gad16}%
  \BibitemOpen
  \bibfield  {author} {\bibinfo {author} {\bibfnamefont {A.~N.}\ \bibnamefont
  {Antonov}}, \bibinfo {author} {\bibfnamefont {M.~K.}\ \bibnamefont
  {Gaidarov}}, \bibinfo {author} {\bibfnamefont {P.}~\bibnamefont
  {Sarriguren}}, \ and\ \bibinfo {author} {\bibfnamefont {E.}~\bibnamefont
  {Moya~de Guerra}},\ }\href {\doibase 10.1103/PhysRevC.94.014319} {\bibfield
  {journal} {\bibinfo  {journal} {Phys. Rev. C}\ }\textbf {\bibinfo {volume}
  {94}},\ \bibinfo {pages} {014319} (\bibinfo {year} {2016})}\BibitemShut
  {NoStop}%
\bibitem [{\citenamefont {Bhuyan}\ \emph {et~al.}(2018)\citenamefont {Bhuyan},
  \citenamefont {Carlson}, \citenamefont {Patra},\ and\ \citenamefont
  {Zhou}}]{bhu18}%
  \BibitemOpen
  \bibfield  {author} {\bibinfo {author} {\bibfnamefont {M.}~\bibnamefont
  {Bhuyan}}, \bibinfo {author} {\bibfnamefont {B.~V.}\ \bibnamefont {Carlson}},
  \bibinfo {author} {\bibfnamefont {S.~K.}\ \bibnamefont {Patra}}, \ and\
  \bibinfo {author} {\bibfnamefont {S.-G.}\ \bibnamefont {Zhou}},\ }\href
  {\doibase 10.1103/PhysRevC.97.024322} {\bibfield  {journal} {\bibinfo
  {journal} {Phys. Rev. C}\ }\textbf {\bibinfo {volume} {97}},\ \bibinfo
  {pages} {024322} (\bibinfo {year} {2018})}\BibitemShut {NoStop}%
\bibitem [{\citenamefont {Kaur}\ \emph
  {et~al.}(2020{\natexlab{a}})\citenamefont {Kaur}, \citenamefont {Quddus},
  \citenamefont {Kumar}, \citenamefont {Bhuyan},\ and\ \citenamefont
  {Patra}}]{kaurnpa}%
  \BibitemOpen
  \bibfield  {author} {\bibinfo {author} {\bibfnamefont {M.}~\bibnamefont
  {Kaur}}, \bibinfo {author} {\bibfnamefont {A.}~\bibnamefont {Quddus}},
  \bibinfo {author} {\bibfnamefont {A.}~\bibnamefont {Kumar}}, \bibinfo
  {author} {\bibfnamefont {M.}~\bibnamefont {Bhuyan}}, \ and\ \bibinfo {author}
  {\bibfnamefont {S.}~\bibnamefont {Patra}},\ }\href {\doibase
  https://doi.org/10.1016/j.nuclphysa.2020.121871} {\bibfield  {journal}
  {\bibinfo  {journal} {Nucl. Phys. A}\ }\textbf {\bibinfo {volume} {1000}},\
  \bibinfo {pages} {121871} (\bibinfo {year} {2020}{\natexlab{a}})}\BibitemShut
  {NoStop}%
\bibitem [{\citenamefont {Kaur}\ \emph
  {et~al.}(2020{\natexlab{b}})\citenamefont {Kaur}, \citenamefont {Quddus},
  \citenamefont {Kumar}, \citenamefont {Bhuyan},\ and\ \citenamefont
  {Patra}}]{kaurjpg}%
  \BibitemOpen
  \bibfield  {author} {\bibinfo {author} {\bibfnamefont {M.}~\bibnamefont
  {Kaur}}, \bibinfo {author} {\bibfnamefont {A.}~\bibnamefont {Quddus}},
  \bibinfo {author} {\bibfnamefont {A.}~\bibnamefont {Kumar}}, \bibinfo
  {author} {\bibfnamefont {M.}~\bibnamefont {Bhuyan}}, \ and\ \bibinfo {author}
  {\bibfnamefont {S.~K.}\ \bibnamefont {Patra}},\ }\href {\doibase
  10.1088/1361-6471/ab92e4} {\bibfield  {journal} {\bibinfo  {journal} {Journal
  of Physics G: Nuclear and Particle Physics}\ }\textbf {\bibinfo {volume}
  {47}},\ \bibinfo {pages} {105102} (\bibinfo {year}
  {2020}{\natexlab{b}})}\BibitemShut {NoStop}%
\bibitem [{\citenamefont {Naik}\ \emph {et~al.}(2019)\citenamefont {Naik},
  \citenamefont {Kaur}, \citenamefont {Kumar},\ and\ \citenamefont
  {Patra}}]{ijmpe}%
  \BibitemOpen
  \bibfield  {author} {\bibinfo {author} {\bibfnamefont {K.~C.}\ \bibnamefont
  {Naik}}, \bibinfo {author} {\bibfnamefont {M.}~\bibnamefont {Kaur}}, \bibinfo
  {author} {\bibfnamefont {A.}~\bibnamefont {Kumar}}, \ and\ \bibinfo {author}
  {\bibfnamefont {S.~K.}\ \bibnamefont {Patra}},\ }\href {\doibase
  10.1142/S0218301319501003} {\bibfield  {journal} {\bibinfo  {journal} {Int.
  J. of Mod. Phys. E}\ }\textbf {\bibinfo {volume} {28}},\ \bibinfo {pages}
  {1950100} (\bibinfo {year} {2019})}\BibitemShut {NoStop}%
\bibitem [{\citenamefont {Quddus}\ \emph {et~al.}(2020)\citenamefont {Quddus},
  \citenamefont {Bhuyan},\ and\ \citenamefont {Patra}}]{abdul19}%
  \BibitemOpen
  \bibfield  {author} {\bibinfo {author} {\bibfnamefont {A.}~\bibnamefont
  {Quddus}}, \bibinfo {author} {\bibfnamefont {M.}~\bibnamefont {Bhuyan}}, \
  and\ \bibinfo {author} {\bibfnamefont {S.~K.}\ \bibnamefont {Patra}},\ }\href
  {\doibase 10.1088/1361-6471/ab4f3e} {\bibfield  {journal} {\bibinfo
  {journal} {Journal of Physics G: Nuclear and Particle Physics}\ }\textbf
  {\bibinfo {volume} {47}},\ \bibinfo {pages} {045105} (\bibinfo {year}
  {2020})}\BibitemShut {NoStop}%
\bibitem [{\citenamefont {Brueckner}\ \emph {et~al.}(1968)\citenamefont
  {Brueckner}, \citenamefont {Buchler}, \citenamefont {Jorna},\ and\
  \citenamefont {Lombard}}]{bruk68}%
  \BibitemOpen
  \bibfield  {author} {\bibinfo {author} {\bibfnamefont {K.~A.}\ \bibnamefont
  {Brueckner}}, \bibinfo {author} {\bibfnamefont {J.~R.}\ \bibnamefont
  {Buchler}}, \bibinfo {author} {\bibfnamefont {S.}~\bibnamefont {Jorna}}, \
  and\ \bibinfo {author} {\bibfnamefont {R.~J.}\ \bibnamefont {Lombard}},\
  }\href {\doibase 10.1103/PhysRev.171.1188} {\bibfield  {journal} {\bibinfo
  {journal} {Phys. Rev.}\ }\textbf {\bibinfo {volume} {171}},\ \bibinfo {pages}
  {1188} (\bibinfo {year} {1968})}\BibitemShut {NoStop}%
\bibitem [{\citenamefont {Brueckner}\ \emph {et~al.}(1969)\citenamefont
  {Brueckner}, \citenamefont {Buchler}, \citenamefont {Clark},\ and\
  \citenamefont {Lombard}}]{bruk69}%
  \BibitemOpen
  \bibfield  {author} {\bibinfo {author} {\bibfnamefont {K.~A.}\ \bibnamefont
  {Brueckner}}, \bibinfo {author} {\bibfnamefont {J.~R.}\ \bibnamefont
  {Buchler}}, \bibinfo {author} {\bibfnamefont {R.~C.}\ \bibnamefont {Clark}},
  \ and\ \bibinfo {author} {\bibfnamefont {R.~J.}\ \bibnamefont {Lombard}},\
  }\href {\doibase 10.1103/PhysRev.181.1543} {\bibfield  {journal} {\bibinfo
  {journal} {Phys. Rev.}\ }\textbf {\bibinfo {volume} {181}},\ \bibinfo {pages}
  {1543} (\bibinfo {year} {1969})}\BibitemShut {NoStop}%
\bibitem [{\citenamefont {Coester}\ \emph {et~al.}(1970)\citenamefont
  {Coester}, \citenamefont {Cohen}, \citenamefont {Day},\ and\ \citenamefont
  {Vincent}}]{coester}%
  \BibitemOpen
  \bibfield  {author} {\bibinfo {author} {\bibfnamefont {F.}~\bibnamefont
  {Coester}}, \bibinfo {author} {\bibfnamefont {S.}~\bibnamefont {Cohen}},
  \bibinfo {author} {\bibfnamefont {B.}~\bibnamefont {Day}}, \ and\ \bibinfo
  {author} {\bibfnamefont {C.~M.}\ \bibnamefont {Vincent}},\ }\href {\doibase
  10.1103/PhysRevC.1.769} {\bibfield  {journal} {\bibinfo  {journal} {Phys.
  Rev. C}\ }\textbf {\bibinfo {volume} {1}},\ \bibinfo {pages} {769} (\bibinfo
  {year} {1970})}\BibitemShut {NoStop}%
\bibitem [{\citenamefont {Brockmann}\ and\ \citenamefont
  {Machleidt}(1990)}]{brockmann}%
  \BibitemOpen
  \bibfield  {author} {\bibinfo {author} {\bibfnamefont {R.}~\bibnamefont
  {Brockmann}}\ and\ \bibinfo {author} {\bibfnamefont {R.}~\bibnamefont
  {Machleidt}},\ }\href {\doibase 10.1103/PhysRevC.42.1965} {\bibfield
  {journal} {\bibinfo  {journal} {Phys. Rev. C}\ }\textbf {\bibinfo {volume}
  {42}},\ \bibinfo {pages} {1965} (\bibinfo {year} {1990})}\BibitemShut
  {NoStop}%
\bibitem [{\citenamefont {Kumar}\ \emph {et~al.}(2018)\citenamefont {Kumar},
  \citenamefont {Patra},\ and\ \citenamefont {Agrawal}}]{kumar18}%
  \BibitemOpen
  \bibfield  {author} {\bibinfo {author} {\bibfnamefont {B.}~\bibnamefont
  {Kumar}}, \bibinfo {author} {\bibfnamefont {S.~K.}\ \bibnamefont {Patra}}, \
  and\ \bibinfo {author} {\bibfnamefont {B.~K.}\ \bibnamefont {Agrawal}},\
  }\href {\doibase 10.1103/PhysRevC.97.045806} {\bibfield  {journal} {\bibinfo
  {journal} {Phys. Rev. C}\ }\textbf {\bibinfo {volume} {97}},\ \bibinfo
  {pages} {045806} (\bibinfo {year} {2018})}\BibitemShut {NoStop}%
\bibitem [{\citenamefont {Kumar}\ \emph {et~al.}(2017)\citenamefont {Kumar},
  \citenamefont {Singh}, \citenamefont {Agrawal},\ and\ \citenamefont
  {Patra}}]{kumar17}%
  \BibitemOpen
  \bibfield  {author} {\bibinfo {author} {\bibfnamefont {B.}~\bibnamefont
  {Kumar}}, \bibinfo {author} {\bibfnamefont {S.}~\bibnamefont {Singh}},
  \bibinfo {author} {\bibfnamefont {B.}~\bibnamefont {Agrawal}}, \ and\
  \bibinfo {author} {\bibfnamefont {S.}~\bibnamefont {Patra}},\ }\href
  {\doibase https://doi.org/10.1016/j.nuclphysa.2017.07.001} {\bibfield
  {journal} {\bibinfo  {journal} {Nuclear Physics A}\ }\textbf {\bibinfo
  {volume} {966}},\ \bibinfo {pages} {197 } (\bibinfo {year}
  {2017})}\BibitemShut {NoStop}%
\bibitem [{\citenamefont {Furnstahl}\ \emph {et~al.}(1996)\citenamefont
  {Furnstahl}, \citenamefont {Serot},\ and\ \citenamefont {Tang}}]{frun96}%
  \BibitemOpen
  \bibfield  {author} {\bibinfo {author} {\bibfnamefont {R.}~\bibnamefont
  {Furnstahl}}, \bibinfo {author} {\bibfnamefont {B.~D.}\ \bibnamefont
  {Serot}}, \ and\ \bibinfo {author} {\bibfnamefont {H.-B.}\ \bibnamefont
  {Tang}},\ }\href {\doibase https://doi.org/10.1016/0375-9474(95)00488-2}
  {\bibfield  {journal} {\bibinfo  {journal} {Nuclear Physics A}\ }\textbf
  {\bibinfo {volume} {598}},\ \bibinfo {pages} {539 } (\bibinfo {year}
  {1996})}\BibitemShut {NoStop}%
\bibitem [{\citenamefont {Furnstahl}\ \emph {et~al.}(1997)\citenamefont
  {Furnstahl}, \citenamefont {Serot},\ and\ \citenamefont {Tang}}]{frun97}%
  \BibitemOpen
  \bibfield  {author} {\bibinfo {author} {\bibfnamefont {R.}~\bibnamefont
  {Furnstahl}}, \bibinfo {author} {\bibfnamefont {B.~D.}\ \bibnamefont
  {Serot}}, \ and\ \bibinfo {author} {\bibfnamefont {H.-B.}\ \bibnamefont
  {Tang}},\ }\href {\doibase https://doi.org/10.1016/S0375-9474(96)00472-1}
  {\bibfield  {journal} {\bibinfo  {journal} {Nuclear Physics A}\ }\textbf
  {\bibinfo {volume} {615}},\ \bibinfo {pages} {441 } (\bibinfo {year}
  {1997})}\BibitemShut {NoStop}%
\bibitem [{\citenamefont {Malik}\ \emph {et~al.}(2018)\citenamefont {Malik},
  \citenamefont {Alam}, \citenamefont {Fortin} \emph {et~al.}}]{malik2018}%
  \BibitemOpen
  \bibfield  {author} {\bibinfo {author} {\bibfnamefont {T.}~\bibnamefont
  {Malik}}, \bibinfo {author} {\bibfnamefont {N.}~\bibnamefont {Alam}},
  \bibinfo {author} {\bibfnamefont {M.}~\bibnamefont {Fortin}},  \emph
  {et~al.},\ }\href {\doibase 10.1103/PhysRevC.98.035804} {\bibfield  {journal}
  {\bibinfo  {journal} {Phys. Rev. C}\ }\textbf {\bibinfo {volume} {98}},\
  \bibinfo {pages} {035804} (\bibinfo {year} {2018})}\BibitemShut {NoStop}%
\bibitem [{\citenamefont {Chen}\ and\ \citenamefont
  {Piekarewicz}(2014)}]{Chen2014}%
  \BibitemOpen
  \bibfield  {author} {\bibinfo {author} {\bibfnamefont {W.-C.}\ \bibnamefont
  {Chen}}\ and\ \bibinfo {author} {\bibfnamefont {J.}~\bibnamefont
  {Piekarewicz}},\ }\href {\doibase 10.1103/PhysRevC.90.044305} {\bibfield
  {journal} {\bibinfo  {journal} {Phys. Rev. C}\ }\textbf {\bibinfo {volume}
  {90}},\ \bibinfo {pages} {044305} (\bibinfo {year} {2014})}\BibitemShut
  {NoStop}%
\bibitem [{\citenamefont {Antonov}\ \emph {et~al.}(1980)\citenamefont
  {Antonov}, \citenamefont {Nikolaev},\ and\ \citenamefont {Petkov}}]{anto1}%
  \BibitemOpen
  \bibfield  {author} {\bibinfo {author} {\bibfnamefont {A.~N.}\ \bibnamefont
  {Antonov}}, \bibinfo {author} {\bibfnamefont {V.~A.}\ \bibnamefont
  {Nikolaev}}, \ and\ \bibinfo {author} {\bibfnamefont {I.~Z.}\ \bibnamefont
  {Petkov}},\ }\href {\doibase 10.1007/BF01892806} {\bibfield  {journal}
  {\bibinfo  {journal} {Z. Phys. A Atoms and Nuclei}\ }\textbf {\bibinfo
  {volume} {297}},\ \bibinfo {pages} {257} (\bibinfo {year}
  {1980})}\BibitemShut {NoStop}%
\bibitem [{\citenamefont {Antonov}\ \emph {et~al.}(1982)\citenamefont
  {Antonov}, \citenamefont {Nikolaev},\ and\ \citenamefont {Petkov}}]{anto2}%
  \BibitemOpen
  \bibfield  {author} {\bibinfo {author} {\bibfnamefont {A.~N.}\ \bibnamefont
  {Antonov}}, \bibinfo {author} {\bibfnamefont {V.~A.}\ \bibnamefont
  {Nikolaev}}, \ and\ \bibinfo {author} {\bibfnamefont {I.~Z.}\ \bibnamefont
  {Petkov}},\ }\href {\doibase 10.1007/BF01414499} {\bibfield  {journal}
  {\bibinfo  {journal} {Z. Phys. A Atoms and Nuclei}\ }\textbf {\bibinfo
  {volume} {304}},\ \bibinfo {pages} {239} (\bibinfo {year}
  {1982})}\BibitemShut {NoStop}%
\bibitem [{\citenamefont {Antonov}\ \emph {et~al.}(1985)\citenamefont
  {Antonov}, \citenamefont {Nikolaev},\ and\ \citenamefont {Petkov}}]{anto3}%
  \BibitemOpen
  \bibfield  {author} {\bibinfo {author} {\bibfnamefont {A.~N.}\ \bibnamefont
  {Antonov}}, \bibinfo {author} {\bibfnamefont {V.~A.}\ \bibnamefont
  {Nikolaev}}, \ and\ \bibinfo {author} {\bibfnamefont {I.~Z.}\ \bibnamefont
  {Petkov}},\ }\href {\doibase 10.1007/BF02905807} {\bibfield  {journal}
  {\bibinfo  {journal} {Nuovo Cimento A}\ }\textbf {\bibinfo {volume} {86}},\
  \bibinfo {pages} {23} (\bibinfo {year} {1985})}\BibitemShut {NoStop}%
\bibitem [{\citenamefont {Antonov}\ \emph {et~al.}(1994)\citenamefont
  {Antonov}, \citenamefont {Kadrev},\ and\ \citenamefont {Hodgson}}]{anto4}%
  \BibitemOpen
  \bibfield  {author} {\bibinfo {author} {\bibfnamefont {A.~N.}\ \bibnamefont
  {Antonov}}, \bibinfo {author} {\bibfnamefont {D.~N.}\ \bibnamefont {Kadrev}},
  \ and\ \bibinfo {author} {\bibfnamefont {P.~E.}\ \bibnamefont {Hodgson}},\
  }\href {\doibase 10.1103/PhysRevC.50.164} {\bibfield  {journal} {\bibinfo
  {journal} {Phys. Rev. C}\ }\textbf {\bibinfo {volume} {50}},\ \bibinfo
  {pages} {164} (\bibinfo {year} {1994})}\BibitemShut {NoStop}%
\bibitem [{\citenamefont {Gaidarov}\ \emph {et~al.}(2011)\citenamefont
  {Gaidarov}, \citenamefont {Antonov}, \citenamefont {Sarriguren},\ and\
  \citenamefont {Moya~de Guerra}}]{gad11}%
  \BibitemOpen
  \bibfield  {author} {\bibinfo {author} {\bibfnamefont {M.~K.}\ \bibnamefont
  {Gaidarov}}, \bibinfo {author} {\bibfnamefont {A.~N.}\ \bibnamefont
  {Antonov}}, \bibinfo {author} {\bibfnamefont {P.}~\bibnamefont {Sarriguren}},
  \ and\ \bibinfo {author} {\bibfnamefont {E.}~\bibnamefont {Moya~de Guerra}},\
  }\href {\doibase 10.1103/PhysRevC.84.034316} {\bibfield  {journal} {\bibinfo
  {journal} {Phys. Rev. C}\ }\textbf {\bibinfo {volume} {84}},\ \bibinfo
  {pages} {034316} (\bibinfo {year} {2011})}\BibitemShut {NoStop}%
\bibitem [{\citenamefont {Gaidarov}\ \emph {et~al.}(2012)\citenamefont
  {Gaidarov}, \citenamefont {Antonov}, \citenamefont {Sarriguren},\ and\
  \citenamefont {de~Guerra}}]{gad12}%
  \BibitemOpen
  \bibfield  {author} {\bibinfo {author} {\bibfnamefont {M.~K.}\ \bibnamefont
  {Gaidarov}}, \bibinfo {author} {\bibfnamefont {A.~N.}\ \bibnamefont
  {Antonov}}, \bibinfo {author} {\bibfnamefont {P.}~\bibnamefont {Sarriguren}},
  \ and\ \bibinfo {author} {\bibfnamefont {E.~M.}\ \bibnamefont {de~Guerra}},\
  }\href {\doibase 10.1103/PhysRevC.85.064319} {\bibfield  {journal} {\bibinfo
  {journal} {Phys. Rev. C}\ }\textbf {\bibinfo {volume} {85}},\ \bibinfo
  {pages} {064319} (\bibinfo {year} {2012})}\BibitemShut {NoStop}%
\bibitem [{\citenamefont {Fuchs}\ \emph {et~al.}(1995)\citenamefont {Fuchs},
  \citenamefont {Lenske},\ and\ \citenamefont {Wolter}}]{fuch95}%
  \BibitemOpen
  \bibfield  {author} {\bibinfo {author} {\bibfnamefont {C.}~\bibnamefont
  {Fuchs}}, \bibinfo {author} {\bibfnamefont {H.}~\bibnamefont {Lenske}}, \
  and\ \bibinfo {author} {\bibfnamefont {H.~H.}\ \bibnamefont {Wolter}},\
  }\href {\doibase 10.1103/PhysRevC.52.3043} {\bibfield  {journal} {\bibinfo
  {journal} {Phys. Rev. C}\ }\textbf {\bibinfo {volume} {52}},\ \bibinfo
  {pages} {3043} (\bibinfo {year} {1995})}\BibitemShut {NoStop}%
\bibitem [{\citenamefont {Antonov}\ \emph {et~al.}(2017)\citenamefont
  {Antonov}, \citenamefont {Kadrev}, \citenamefont {Gaidarov}, \citenamefont
  {Sarriguren},\ and\ \citenamefont {de~Guerra}}]{anto17}%
  \BibitemOpen
  \bibfield  {author} {\bibinfo {author} {\bibfnamefont {A.~N.}\ \bibnamefont
  {Antonov}}, \bibinfo {author} {\bibfnamefont {D.~N.}\ \bibnamefont {Kadrev}},
  \bibinfo {author} {\bibfnamefont {M.~K.}\ \bibnamefont {Gaidarov}}, \bibinfo
  {author} {\bibfnamefont {P.}~\bibnamefont {Sarriguren}}, \ and\ \bibinfo
  {author} {\bibfnamefont {E.~M.}\ \bibnamefont {de~Guerra}},\ }\href {\doibase
  10.1103/PhysRevC.95.024314} {\bibfield  {journal} {\bibinfo  {journal} {Phys.
  Rev. C}\ }\textbf {\bibinfo {volume} {95}},\ \bibinfo {pages} {024314}
  (\bibinfo {year} {2017})}\BibitemShut {NoStop}%
\bibitem [{\citenamefont {Fujita}\ and\ \citenamefont
  {Miyazawa}(1957)}]{fujita}%
  \BibitemOpen
  \bibfield  {author} {\bibinfo {author} {\bibfnamefont {J.-i.}\ \bibnamefont
  {Fujita}}\ and\ \bibinfo {author} {\bibfnamefont {H.}~\bibnamefont
  {Miyazawa}},\ }\href {\doibase 10.1143/PTP.17.360} {\bibfield  {journal}
  {\bibinfo  {journal} {Progress of Theoretical Physics}\ }\textbf {\bibinfo
  {volume} {17}},\ \bibinfo {pages} {360} (\bibinfo {year} {1957})}\BibitemShut
  {NoStop}%
\bibitem [{\citenamefont {Pieper}\ \emph {et~al.}(2001)\citenamefont {Pieper},
  \citenamefont {Pandharipande}, \citenamefont {Wiringa},\ and\ \citenamefont
  {Carlson}}]{Pandhere}%
  \BibitemOpen
  \bibfield  {author} {\bibinfo {author} {\bibfnamefont {S.~C.}\ \bibnamefont
  {Pieper}}, \bibinfo {author} {\bibfnamefont {V.~R.}\ \bibnamefont
  {Pandharipande}}, \bibinfo {author} {\bibfnamefont {R.~B.}\ \bibnamefont
  {Wiringa}}, \ and\ \bibinfo {author} {\bibfnamefont {J.}~\bibnamefont
  {Carlson}},\ }\href {\doibase 10.1103/PhysRevC.64.014001} {\bibfield
  {journal} {\bibinfo  {journal} {Phys. Rev. C}\ }\textbf {\bibinfo {volume}
  {64}},\ \bibinfo {pages} {014001} (\bibinfo {year} {2001})}\BibitemShut
  {NoStop}%
\bibitem [{\citenamefont {{De Vries}}\ \emph {et~al.}(1987)\citenamefont {{De
  Vries}}, \citenamefont {{De Jager}},\ and\ \citenamefont {{De
  Vries}}}]{varies87}%
  \BibitemOpen
  \bibfield  {author} {\bibinfo {author} {\bibfnamefont {H.}~\bibnamefont {{De
  Vries}}}, \bibinfo {author} {\bibfnamefont {C.~W.}\ \bibnamefont {{De
  Jager}}}, \ and\ \bibinfo {author} {\bibfnamefont {C.}~\bibnamefont {{De
  Vries}}},\ }\href {\doibase https://doi.org/10.1016/0092-640X(87)90013-1}
  {\bibfield  {journal} {\bibinfo  {journal} {Atomic Data and Nucl. Data
  Tables}\ }\textbf {\bibinfo {volume} {36}},\ \bibinfo {pages} {495 }
  (\bibinfo {year} {1987})}\BibitemShut {NoStop}%
\bibitem [{\citenamefont {Nadjakov}\ \emph {et~al.}(1994)\citenamefont
  {Nadjakov}, \citenamefont {Marinova},\ and\ \citenamefont
  {Gangrsky}}]{nadj94}%
  \BibitemOpen
  \bibfield  {author} {\bibinfo {author} {\bibfnamefont {E.}~\bibnamefont
  {Nadjakov}}, \bibinfo {author} {\bibfnamefont {K.}~\bibnamefont {Marinova}},
  \ and\ \bibinfo {author} {\bibfnamefont {Y.}~\bibnamefont {Gangrsky}},\
  }\href {\doibase https://doi.org/10.1006/adnd.1994.1004} {\bibfield
  {journal} {\bibinfo  {journal} {Atomic Data and Nucl. Data Tables}\ }\textbf
  {\bibinfo {volume} {56}},\ \bibinfo {pages} {133 } (\bibinfo {year}
  {1994})}\BibitemShut {NoStop}%
\bibitem [{\citenamefont {Danielewicz}(2003)}]{dani03}%
  \BibitemOpen
  \bibfield  {author} {\bibinfo {author} {\bibfnamefont {P.}~\bibnamefont
  {Danielewicz}},\ }\href {\doibase
  https://doi.org/10.1016/j.nuclphysa.2003.08.001} {\bibfield  {journal}
  {\bibinfo  {journal} {Nucl. Phys. A}\ }\textbf {\bibinfo {volume} {727}},\
  \bibinfo {pages} {233 } (\bibinfo {year} {2003})}\BibitemShut {NoStop}%
\bibitem [{\citenamefont {Danielewicz}(2004)}]{dani04}%
  \BibitemOpen
  \bibfield  {author} {\bibinfo {author} {\bibfnamefont {P.}~\bibnamefont
  {Danielewicz}},\ }\href@noop {} {\  (\bibinfo {year} {2004})},\ \Eprint
  {http://arxiv.org/abs/0411115} {arXiv:0411115} \BibitemShut {NoStop}%
\bibitem [{\citenamefont {Danielewicz}(2007)}]{dani06}%
  \BibitemOpen
  \bibfield  {author} {\bibinfo {author} {\bibfnamefont {P.}~\bibnamefont
  {Danielewicz}},\ }\enquote {\bibinfo {title} {Symmetry energy},}\ in\ \href
  {\doibase 10.1142/9789812708250_0015} {\emph {\bibinfo {booktitle}
  {Opportunities with Exotic Beams}}}\ (\bibinfo {year} {2007})\ pp.\ \bibinfo
  {pages} {142--151}\BibitemShut {NoStop}%
\bibitem [{\citenamefont {{ A. E. L.}Dieperink}\ and\ \citenamefont {{P. Van
  Isacker}}(2007)}]{diep07}%
  \BibitemOpen
  \bibfield  {author} {\bibinfo {author} {\bibnamefont {{ A. E. L.}Dieperink}}\
  and\ \bibinfo {author} {\bibnamefont {{P. Van Isacker}}},\ }\href {\doibase
  10.1140/epja/i2007-10360-3} {\bibfield  {journal} {\bibinfo  {journal} {Eur.
  Phys. J. A}\ }\textbf {\bibinfo {volume} {32}},\ \bibinfo {pages} {11}
  (\bibinfo {year} {2007})}\BibitemShut {NoStop}%
\bibitem [{\citenamefont {Danielewicz}\ and\ \citenamefont
  {Lee}(2009)}]{dani09}%
  \BibitemOpen
  \bibfield  {author} {\bibinfo {author} {\bibfnamefont {P.}~\bibnamefont
  {Danielewicz}}\ and\ \bibinfo {author} {\bibfnamefont {J.}~\bibnamefont
  {Lee}},\ }\href {\doibase 10.1142/S0218301309013014} {\bibfield  {journal}
  {\bibinfo  {journal} {Int. J. of Mod. Phys. E}\ }\textbf {\bibinfo {volume}
  {18}},\ \bibinfo {pages} {892} (\bibinfo {year} {2009})}\BibitemShut
  {NoStop}%
\bibitem [{\citenamefont {Antonov}\ \emph {et~al.}(2018)\citenamefont
  {Antonov}, \citenamefont {Kadrev}, \citenamefont {Gaidarov}, \citenamefont
  {Sarriguren},\ and\ \citenamefont {Moya~de Guerra}}]{anto18}%
  \BibitemOpen
  \bibfield  {author} {\bibinfo {author} {\bibfnamefont {A.~N.}\ \bibnamefont
  {Antonov}}, \bibinfo {author} {\bibfnamefont {D.~N.}\ \bibnamefont {Kadrev}},
  \bibinfo {author} {\bibfnamefont {M.~K.}\ \bibnamefont {Gaidarov}}, \bibinfo
  {author} {\bibfnamefont {P.}~\bibnamefont {Sarriguren}}, \ and\ \bibinfo
  {author} {\bibfnamefont {E.}~\bibnamefont {Moya~de Guerra}},\ }\href
  {\doibase 10.1103/PhysRevC.98.054315} {\bibfield  {journal} {\bibinfo
  {journal} {Phys. Rev. C}\ }\textbf {\bibinfo {volume} {98}},\ \bibinfo
  {pages} {054315} (\bibinfo {year} {2018})}\BibitemShut {NoStop}%
\bibitem [{\citenamefont {Bhuyan}\ \emph {et~al.}(2020)\citenamefont {Bhuyan},
  \citenamefont {Kumar}, \citenamefont {Rana}, \citenamefont {Jain},
  \citenamefont {Patra},\ and\ \citenamefont {Carlson}}]{bhu20}%
  \BibitemOpen
  \bibfield  {author} {\bibinfo {author} {\bibfnamefont {M.}~\bibnamefont
  {Bhuyan}}, \bibinfo {author} {\bibfnamefont {R.}~\bibnamefont {Kumar}},
  \bibinfo {author} {\bibfnamefont {S.}~\bibnamefont {Rana}}, \bibinfo {author}
  {\bibfnamefont {D.}~\bibnamefont {Jain}}, \bibinfo {author} {\bibfnamefont
  {S.~K.}\ \bibnamefont {Patra}}, \ and\ \bibinfo {author} {\bibfnamefont
  {B.~V.}\ \bibnamefont {Carlson}},\ }\href {\doibase
  10.1103/PhysRevC.101.044603} {\bibfield  {journal} {\bibinfo  {journal}
  {Phys. Rev. C}\ }\textbf {\bibinfo {volume} {101}},\ \bibinfo {pages}
  {044603} (\bibinfo {year} {2020})}\BibitemShut {NoStop}%
\end{thebibliography}%
\bibliographystyle{apsrev4-1}
%%%%
\begin{comment}

%%
\end{comment}
\end{document}